\documentclass[sigplan,nonacm]{acmart}
\usepackage{subcaption} 
\usepackage{graphicx}
\usepackage{booktabs}
\usepackage{tabularx}
\usepackage{threeparttable}
\usepackage{multirow}
\usepackage{mathtools}
\usepackage{amsthm}
\usepackage{amsmath,amsfonts}
\usepackage[noend]{algpseudocode}
\usepackage[vlined,linesnumbered,ruled]{algorithm2e}
\makeatletter
\renewcommand{\tcp}[1]{\textcolor{blue}{\small\(\triangleright\)~#1}\par}
\usepackage{tikz}
\usepackage{balance}
\usepackage[capitalize, noabbrev]{cleveref}
\usepackage{hyperref}

\newcommand{\argmax}{\operatorname*{arg\,max}}
\newcommand{\argmin}{\operatorname*{arg\,min}}
\def\ie{\textit{i.e.}\xspace}
\def\vs{\textit{vs.}\xspace}

\def\eg{\textit{e.g.}\xspace}

\def\st{\xspace\textbf{s.t.}\xspace}

\begin{document}

% CoSine: Accelerating Large Language Model Serving with Collaborative Multi-node Speculation  Collaborative Speculative Inference in Heterogeneous Environments
% via Exploiting GPU Heterogeneity Collaborative Speculative Inference for Efficient LLM Inference Serving
\title[CoSine]{Collaborative Speculative Inference for
Efficient LLM Inference Serving}
\author{Luyao Gao, Jianchun Liu, Hongli Xu, Xichong Zhang, Yunming Liao, Liusheng Huang}
\affiliation{
\institution{University of Science and Technology of China}
\city{Hefei}
\country{China}
}
% across multi-node collaboration with heterogeneous resources.  Collaborative Speculative Inference with Multi-node Heterogeneous Resources
% xxx-guide / xxx-based / xxx-driven 
% optimized exploration / xxx-enabled 

\begin{abstract}
%Abstract
% Large language models (LLMs) have demonstrated remarkable capabilities in inference tasks across diverse applications.
Speculative inference is a promising paradigm employing small speculative models (SSMs) as drafters to generate draft tokens, which are subsequently verified in parallel by the target large language model (LLM).
This approach enhances the efficiency of inference serving by reducing LLM inference latency and costs while preserving generation quality.
% challenges
However, existing speculative methods face critical challenges, including inefficient resource utilization and limited draft acceptance, which constrain their scalability and overall effectiveness.
% , due to 
% introduce CoSine decouple and collaboration
%To overcome these obstacles, we present \textit{CoSine}, a novel collaborative speculative inference system across multi-nodes with heterogeneous GPU resources.
To overcome these obstacles, we present \textit{CoSine}, a novel speculative inference system that decouples sequential speculative decoding from parallel verification, enabling efficient collaboration among multiple nodes. %to mitigate resource bottlenecks and improve draft quality. 
%
%With the collaboration of multiple drafters, 
Specifically, CoSine routes inference requests to specialized drafters based on their expertise and incorporates a confidence-based token fusion mechanism to synthesize outputs from cooperating drafters, ensuring high-quality draft generation.
Additionally, CoSine dynamically orchestrates the execution of speculative decoding and verification in a pipelined manner, employing batch scheduling to selectively group requests and adaptive speculation control to minimize idle periods. 
By optimizing parallel workflows through heterogeneous node collaboration, CoSine balances draft generation and verification throughput in real time, thereby maximizing resource utilization.
% implements adaptive resource management that balances GPU resource contention
Experimental results demonstrate that CoSine achieves superior performance compared to state-of-the-art speculative approaches.
Notably, with equivalent resource costs, CoSine achieves up to a 23.2\% decrease in latency and a 32.5\% increase in throughput compared to baseline methods.

\end{abstract}


\maketitle

\pagestyle{plain}

\section{Introduction} \label{sec:intro}

Recent advancements in large language models (LLMs) have showcased remarkable capabilities in language understanding and generation, as well as adaptability across diverse academic and industrial domains \cite{vaswani2017attention}. 
Among the various architectures, decoder-only Transformer models, such as the GPT series \cite{achiam2023gpt} and the Llama family \cite{touvron2023llama}, have emerged as highly prominent due to their simplicity and effectiveness. 
These models employ uniform decoder layers within an autoregressive framework, generating text iteratively by predicting subsequent tokens with previous tokens.
The autoregressive paradigm is particularly well-suited for applications requiring high accuracy and contextual comprehension, including virtual assistants, and program synthesis \cite{yuan2024llm}.
%These models typically employ uniform decoder layers in an autoregressive paradigm, enabling iterative text generation by predicting subsequent tokens based on previously generated tokens. , and search engines

However, deploying LLM inference services efficiently and cost-effectively remains challenging, due to the escalating parameter size and increasing architectural complexity \cite{xu2024survey}.
For instance, the Mixture-of-Experts model DeepSeek-R1, with its 685 billion parameters, requires over 1,500 GB GPU memory to perform inference \cite{guo2025deepseek}.
Many existing LLM systems rely on {incremental decoding}, a process that generates one token at a time while re-computing the activations of all preceding tokens. 
The massive parameters introduce substantial computational and memory overheads, incurring significant serving costs in real-time deployments (\eg, OpenAI o1 model costs \$60 per 1M output tokens) \cite{achiam2023gpt}.

Inspired by branch prediction techniques in CPU architectures \cite{chen2023accelerating}, \textit{speculative inference} \cite{leviathan2023fast} has been introduced to mitigate inference overhead without compromising generation quality. 
% Speculative decoding leverages token-level variability in computational demands by combining large and small models to approximate the performance of the large model while mitigating its high computational overhead. % accept a subset of them at continuous positions 
This approach allows small speculative models (SSMs) as drafters to generate successive draft tokens through autoregressive decoding, while the LLM verifies these tokens in parallel using rejection sampling \cite{xu2024survey}.
% with batch processing approach for the reuse of layer weights
Because SSMs incur lower computational costs, many draft tokens can be accepted under the same probability distribution without requiring iterative decoding in the LLM, leading to substantial speedup \cite{xia2024unlocking}. 
Furthermore, speculative inference does not necessitate additional retraining or fine-tuning of the LLM and can be readily integrated into well-established acceleration frameworks like vLLM \cite{yuan2024llm}.
% and HuggingFace Transformers \cite{wolf2020transformers}

Despite its promise, speculative inference faces two major challenges that hinder widespread adoption. 
(1) \textbf{Disparate Resource Demands:} % bound Disparate Demands computational asymmetry 
Speculative inference divergent resource demands stemming from the architectural disparity between memory-intensive SSMs and compute-intensive LLMs \cite{miao2023towards}.
Although SSMs incur roughly $1,000\times$ lower computation overhead compared to LLMs, their operation necessitates sustained high memory bandwidth for efficient token generation  (see detailed in Section~\ref{sec:motivation}). Co-location of speculative drafting and verification phases within shared compute pools exacerbates resource contention, particularly when models and key-value caches maintained simultaneously \cite{chen2024sequoia}. % SSMs and LLMs interfere idle periods and resource contention%
Furthermore, both datacenter GPUs (\eg, NVIDIA A100) and consumer GPUs (\eg, RTX 2080Ti) demonstrate suboptimal utilization when simultaneously processing compute bound LLMs and memory bound SSMs, leading to resource contention and elevated costs \cite{liu2024optimizing}. 
(2) \textbf{Constrained Drafter Knowledge:} 
As discussed in Section~\ref{sec:motivation}, drafters often struggle to generate high-quality draft tokens that pass verification at acceptable rates (\eg, the LLaMA13B-LLaMA68M \ model pair exhibits acceptance rates below 0.3) \cite{wan2024knowledge}.
Efforts to improve acceptance by expanding the length or breadth of draft tokens often yield diminishing returns, primarily due to the limited generalization capabilities of drafters \cite{wang2024minions}. 
This limitation becomes more pronounced in complex tasks with longer sequences, where drafters fail to generate sufficiently accurate drafts, undermining the speedup potential of speculative inference \cite{yang2024multi}.
%

% current related work, add the reason why they are not enough, current sentence is too weak and not clear
Existing works primarily focus on improving speculative inference by enhancing draft acceptance and resource utilization.
First, some studies \cite{cai2024medusa, fu2024break, li2024eagle2} aim to adopt specialized token-tree structures to increase draft quality and acceptance, but generally fail to strike a balance between computational overhead and draft quality.
For example, OPT-Tree \cite{wang2024opt} and EAGLE2 \cite{li2024eagle2} leverage mathematical optimizing and context-aware fine-tuning to explore draft structures and relationships.
% draft structures and acceptance.
However, they demand substantial resources for probility and distribution caculation, limiting their scalability and practical adoption in resource-constrained scenarios.
Second, other researches \cite{yang2024multi, butler2024pipeinfer, santilli2023accelerating} focus on inference parallelism and batched processing to enhance resource utilization, while frequently cannot adapt to dynamic inference workload or verification status.
For instance, PipeInfer \cite{butler2024pipeinfer} executes decoding and verification pipelines in parallel but cannot dynamically adapt resource allocation between drafting and verification based on runtime conditions, leading to suboptimal speculative efficiency with varying workloads.
% lacks resource reallocate mechanisms based on run-time conditions, leading to inefficient draft generation. 

Such inefficiencies primarily arise from the sequential execution of draft generation coupled with parallel execution of verification in speculative inference.
Fortunately, the token-level exchange in speculative inference enables the decoupling of these two phases across multiple nodes, such as single-GPU devices and multi-GPU servers \cite{lim2024accelerating}. 
This decoupled approach permits independent resource allocation and adaptive workflow scheduling, thereby improving resource utilization and system scalability \cite{chen2024sequoia}. 
However, concerns regarding increased operational costs due to hardware scaling may arise, as parallel verification demands computational capability while draft generation prioritizes memory bandwidth \cite{yuan2024llm}. %GPU
Major cloud providers (\eg, AWS and Azure) offer various GPUs that support multiple node collaboration with heterogeneous resources \cite{xia2024unlocking}. % and Google Cloud
In fact, industry and academic communities have increasingly adopted mixed GPU clusters composed of both high-performance and cost-effective nodes for inference \cite{lim2024accelerating}. 
% summary and combine with following
Consequently, leveraging heterogeneous GPU resources and multiple nodes collaboration is essential to the decoupled speculative inference \cite{chen2024sequoia}. %enhance inference performance 

%  On the other hand, one may think the decoupling method will increase operational costs due to hardware scaling [7]. As we know, two phases require different resource consumptions. Major cloud providers (XXX) offer various GPU than ...... In fact, industry and academic communities have increasingly adopted mixed GPU clusters  for inference [16].  

% CoSine introduction
To this end, we present \textit{CoSine}, a novel speculative inference system to facilitate collaboration among multiple nodes, enabling efficient and cost-effective LLM serving with heterogeneous GPU resources.
Specifically, CoSine decouples sequential speculative decoding from parallel verification, enabling efficient collaboration among multiple nodes and adaptively assigning the most suitable resources to each phase.
With multiple drafters collaborating to generate drafts in parallel, CoSine dynamically routes inference requests to optimally suited drafters, leveraging their specialized expertise across different domains.
CoSine further introduces a confidence-based token fusion mechanism that synthesizes outputs across multiple drafters, enabling high-quality draft generation with collective expertise.
Additionally, CoSine dynamically orchestrates speculative decoding and verification phases in a pipelined manner, employing batch scheduling to selectively group requests and adaptive speculation control to minimize pipeline idle periods.
By optimizing parallel workflows through heterogeneous node collaboration, CoSine balances draft generation and verification throughput in real time, maximizing resource utilization.
% Additionally, we orchestrate the execution of decoding and verification in the pipeline, optimizing parallel workflows with heterogeneous node collaboration to balance draft generation and verification throughput in real time.
% % minimize idle time and maximize resource utilization.  separately address their distinct demands.
% CoSine dynamically allocates resources for drafters and LLM, separately addressing their distinct demands in the pipeline
% continuously balancing draft quality and verification throughput. %pipeline  to achieve optimal performance
% perceiving dynamic resource demands and verification status in real-time, minimizing idle time and optimizing system throughput. request batch optimization

%CoSine solved challenges  in optimizing speculative inference
However, CoSine confronts two fundamental challenges. 
First, while domain-specific drafters enable targeted generation, their constrained parameters introduce inherent sensitivity to cross-domain requests \cite{timor2024distributed}. 
The increasing diversity of request patterns demands real-time adaptability in routing strategy to maintain generation quality.
Assigning an improper drafter to a request may cause significant degradation in draft acceptance, necessitating an \textit{adaptive request routing strategy} based on drafter expertise profiles and historical verification patterns.
% collaboration with heterogeneous resources creates diverse execution behaviors in iterative workflows. 
Second, the temporal unbalance between draft generation and verification presents coordination challenges \cite{butler2024pipeinfer}. 
While verification latency remains predictable through fixed parallel execution of large models, sequential draft generation exhibits significant variance across requests \cite{bhendawade2024speculative}. 
Uncoordinated pipeline management risks frequent stalls and high rejection rates, undermining speculative inference benefits. 
Hence, CoSine needs to \textit{balance the draft generation and verification in pipeline workflows} through the continuous perception of resource demands and request workload, ensuring minimizing idle time during LLM serving.
% and optimizing resource utilization.
We summarize our contributions as follows:

\begin{itemize}
    \item We present \textit{CoSine}, a novel speculative inference system for architectural decoupling of sequential speculative decoding and parallel verification phases with multi-node collaboration. This design improves both draft generation efficiency and heterogeneous resource utilization in LLM inference serving.
    \item CoSine employs an adaptive strategy to route inference requests to the suitable drafters, and further enhances draft quality through a confidence-based token fusion mechanism that synthesizes outputs across multiple nodes. 
    \item CoSine dynamically orchestrates pipeline workflows by balancing resource allocation between draft generation and verification in real time, optimizing both acceptance rates and resource utilization.  %through demand and verification status perception
    \item Extensive experiments demonstrate that CoSine outperforms state-of-the-art speculative inference systems. Notably, under equivalent costs, CoSine achieves up to a 23\% reduction in latency and a 32\% improvement in throughput compared to baseline methods. 
\end{itemize}

%that achieves latency reduction and throughput enhancement 

% draft generation often fluctuates and verification relatively stable
% Through continuous monitoring of verification states and dynamic workload balancing

\section{Background}\label{sec:bkground} 
Transformer-based LLMs, such as LLaMA \cite{touvron2023llama} and DeepSeek \cite{deepseek-llm}, operate through two primary phases: the {prefill} phase and the {decoding} phase.
In the prefill phase, LLM processes input prompt tokens in parallel, constructing a key-value (KV) cache to capture inter-token relationships. 
The decoding phase then generates tokens sequentially, with each new token depending on the KV cache and preceding tokens.
In this work, we focus on speculative inference and model ensemble to enhance the efficiency of multi-node collaboration.

% across multi-node collaboration with heterogeneous resources. 

\subsection{Speculative Inference} 
\begin{figure}[t]
    \centering
    \includegraphics[width=0.92\linewidth]{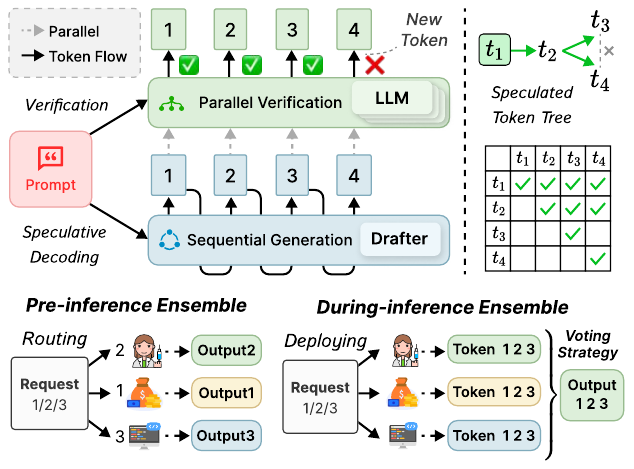}
    % \hfill
    % \begin{subfigure}[b]{0.48\columnwidth}
    %     \includegraphics[width=\linewidth]{chap2/pic/ensemble.pdf}
    %     \caption{LLM ensemble.}
    %     \label{fig:b2}
    % \end{subfigure}
    \caption{Overview of speculative inference with verification and speculative decoding phases. Besides, we present the LLM ensemble with pre-inference and during-inference.}
    \label{fig:b12}
\end{figure}

Speculative inference accelerates LLM inference while preserving generation quality according to the observation that many easy tokens can be predicted with less computational overhead \cite{chen2023accelerating}.
It has a two-phase process: speculative decoding and parallel verification, as illustrated in \cref{fig:b12}.
The speculative decoding phase employs SSMs as the {drafter} to autoregressively generate draft tokens, utilizing same tokenizer as the target LLM \cite{yuan2024llm}.
The verification phase then processes these tokens in parallel using the target LLM, leveraging batched weight reuse and reduced memory access overhead \cite{leviathan2023fast}. 
To maintain distributional alignment with the target LLM, an acceptance-rejection mechanism ensures that accepted tokens match the distribution of target LLM \cite{liu2024optimizing}.

% Formally, during each iteration, the drafter generates a candidate token sequence \((x_1, x_2, \dots, x_s)\), and the LLM computes the corresponding logits \(o_i(x)\) in parallel. 
% These logits are compared against the drafter’s logits \(q_i(x)\), with tokens accepted if \(q_i(x) \le o_i(x)\) or rejected with probability \(1 - \tfrac{o_i(x)}{q_i(x)}\).
% Rejected tokens are resampled from an adjusted distribution \(\mathrm{norm}(\max(0,\, o_i(x) - q_i(x)))\), with a single rejection discarding all subsequent tokens in the speculation phase.
% If all tokens are accepted, the target LLM samples an additional token from \(o_{i+1}(x)\), iterating until sequence completion.
% %
Formally, the drafter generates $\gamma$ candidate tokens $\mathbf{X}_\text{draft} = (x_1, \ldots, x_\gamma)$, where $x_i \sim q(\cdot \mid \mathbf{x}_{<i})$. The target LLM computes logits $o_i(x) = o(x \mid \mathbf{x}_{<i})$ for each token.
The acceptance-rejection mechanism compares the drafter's logits $q_i(x)$ with the target LLM's logits $o_i(x)$, accepting tokens if $q_i(x) \le o_i(x)$ or rejecting them with probability $1 - \tfrac{o_i(x)}{q_i(x)}$.
Rejected tokens are resampled from a distribution $\mathrm{norm}(\max\{0,\, o_i(x) - q_i(x)\})$, with a single rejection discarding all subsequent tokens in the speculation phase.
If all tokens are accepted, the target LLM samples an additional token from $x_{\gamma+1} \sim o(\cdot \mid \mathbf{X}_\text{draft})$.  
To enhance acceptance rates, multiple drafters can generate parallel draft sequences and merge them into a tree topology that preserves causal dependencies based on tree attention \cite{wang2024opt}. 
% Rather than computing this tree topology through computationally expensive depth-first traversals, recent research employs causal masking based on tree attention, enabling parallel verification of all draft sequences \cite{miao2024specinfer}.

Furthermore, self-speculation methods like lookahead decoding \cite{fu2024break} and Medusa \cite{cai2024medusa} extend the principles of parallel decoding and speculative inference by incorporating future token considerations into the decoding process without requiring additional fine-tuning or draft models. 
Unlike traditional decoding methods, which make irrevocable token choices at each step, lookahead decoding predicts N-grams directly from token probabilities in parallel, generating multiple draft tokens to identify the optimal subsequence. 
Medusa \cite{cai2024medusa} adds new sampling heads to the target LLM that produce the speculations without a speculative model, requiring training new sampling heads for the target LLM. %
These approaches leverage the model's inherent capability to predict sequences, allowing it to avoid local optima and address the limitations of greedy decoding strategies \cite{wan2024knowledge}.

\subsection{Large Language Model Ensemble}
Reliance on single LLM for critical generative tasks such as essay composition and code synthesis exposes inherent limitations, including output inconsistencies, stylistic biases, and inadequate domain adaptation \cite{lu2023routing, li2024more}. 
The growing diversity of open-source LLMs has catalyzed advancements in ensemble methods that strategically combine multiple pretrained models to enhance output quality while compensating for individual weaknesses, as shown in \cref{fig:b12}.
Such ensembles employ knowledge fusion techniques to cross-validate information across models, thereby reducing factual inaccuracies and hallucinations \cite{wan2024knowledge}. 
For instance, essay composition demands logical structure, evidence integration, and domain-specific knowledge, requiring coordinated ensemble approaches to enhance argumentative depth \cite{xia2024unlocking}.
% By harmonizing diverse outputs, ensembles enhance argumentative depth and stylistic consistency. 
In this work, we focus on pre-inference ensemble and during-inference ensemble to coordinate multiple models for the quality and adaptive LLM generation.

Pre-inference ensembles optimize task performance by matching each request to the suitable models through real-time analysis of input features \cite{li2024more}.
It utilize intelligent request routing strategies, such as domain-specific routing, computational constraints, or model specialization.
During-inference ensembles operate through real-time integration of probability distributions or logits from multiple LLMs during token generation \cite{fu2025speculative}.
This can be achieved through weighted averaging of distributions, contrastive decoding to amplify differences between candidate outputs, or trainable mechanisms that adaptively combine model outputs \cite{lu2023routing}. 
Through strategic synthesis of diverse linguistic patterns and knowledge representations, these techniques significantly improve the adaptability and robustness of LLM generation.

\section{Motivation}\label{sec:motivation}

% bound/bottleneck

\begin{figure}[t]
    \centering
    \begin{subfigure}[b]{0.49\columnwidth}
        \includegraphics[width=\linewidth]{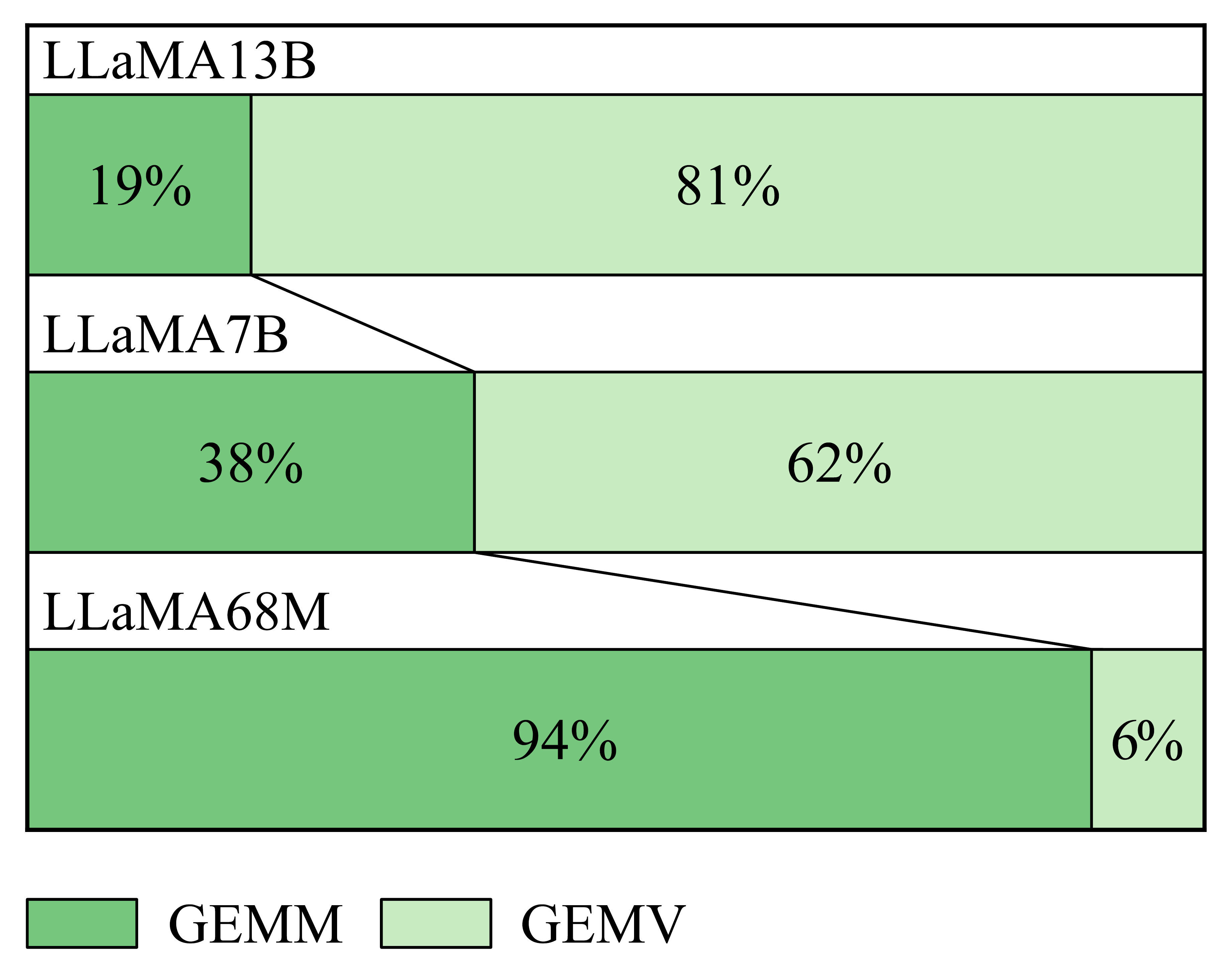}
        \caption{Latency proportion of GEMM and GEMV in LLM inference operators.}
        \label{fig:m1}
    \end{subfigure}
    % \hfill
    \begin{subfigure}[b]{0.49\columnwidth}
        \includegraphics[width=\linewidth]{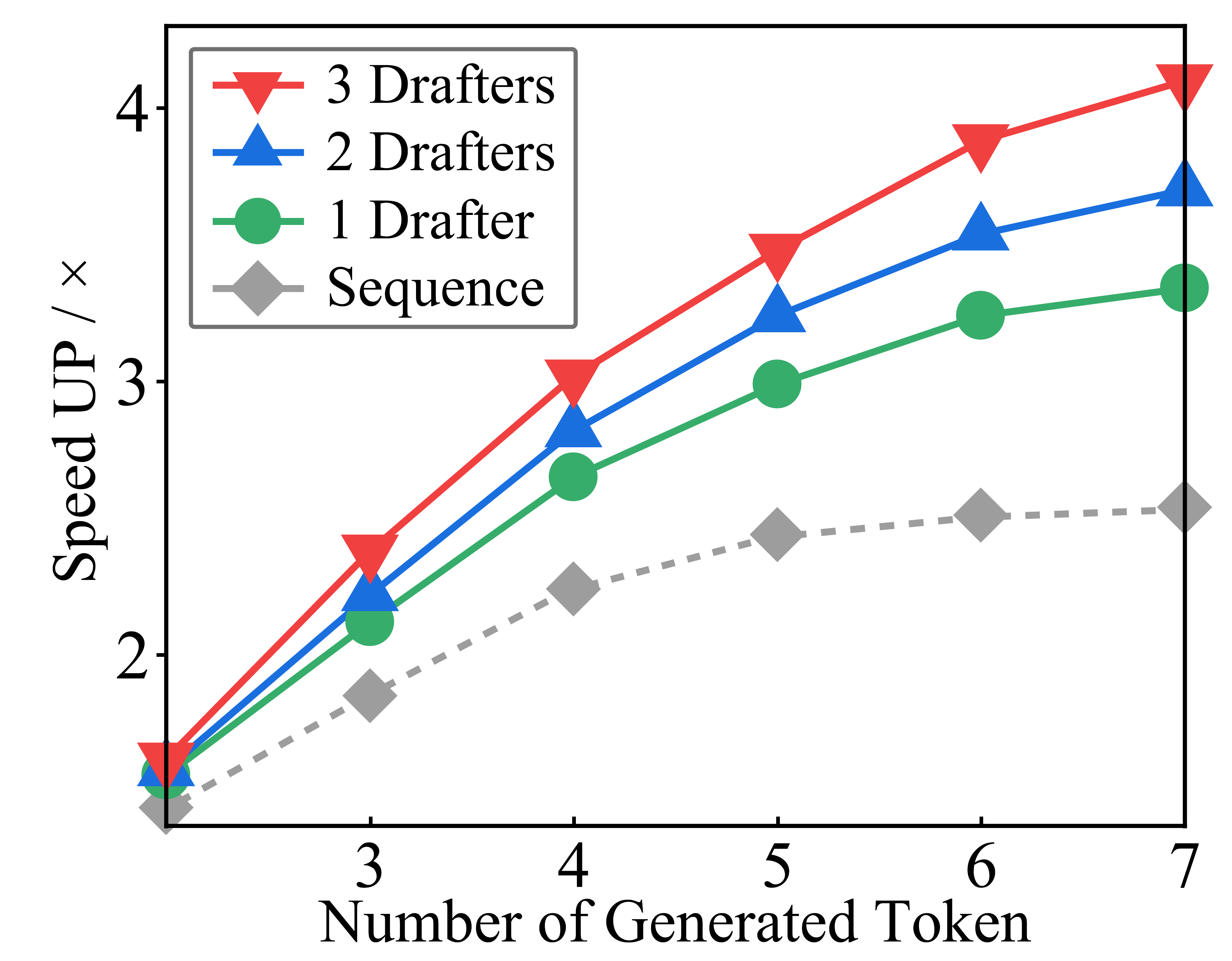}
        \caption{Inference speedup across different structures and drafter numbers.}
        \label{fig:m2}
    \end{subfigure}
    \caption{Performance bounds across model architectures and drafting configurations.}
    \label{fig:m12}
\end{figure}
        
\subsection{Key Observations in Speculative Inference}

First, current speculative inference systems exhibit a fundamental mismatch between the alternating demands of speculative decoding and verification phases.
%
% Two fundamental operations that play a critical role in LLM forward propagation are GEMM (General Matrix-Matrix Multiplication) and GEMV (General Matrix-Vector Multiplication), which are compute-intensive and memory-intensive, respectively.
% require powerful computing capability and memory access
To analyze these bottlenecks, we profile the proportion of GEMM (General Matrix-Matrix Multiplication) and GEMV (General Matrix-Vector Multiplication) operations during SSM-based sequential drafting and LLM-based parallel verification on a NVIDIA A100 GPU, as shown in \cref{fig:m1}.
The autoregressive speculative drafting process in SSMs relies predominantly on memory-intensive GEMV operations, requiring rapid access to token embeddings and weight matrices.
In contrast, batched verification in LLMs emphasizes compute-intensive GEMM operations optimized for parallel computations. 

Conventional shared-resource execution either remains computational units underutilized during drafting or memory bandwidth idle during verification.
Our observation is that coupled designs for speculative inference struggle to simultaneously satisfy these divergent needs, due to the inherent mismatch between GEMM and GEMV dominated operations.
Thereby, a new opportunity arises to decouple these phases and leverage heterogeneous resources to optimize each phase independently.

Second, simply increasing draft length to improve token acceptance proves suboptimal due to diminishing returns in inference acceleration. 
To evaluate drafting strategy impacts on generation quality, we measure speculative inference speedup across varying draft structures using multiple LLaMA-68M for drafting and LLaMA-13B for verification, as detailed in \cref{fig:m2}.
It reveal that sequential drafting exhibits progressively smaller speedup gains as length increase, while tree-structured drafts expand the candidate space to better align with LLM token distribution.
Furthermore, aggregating predictions from multiple drafters achieves greater coverage of the probabilistic space of LLM to speedup inference.
The observation is that multi-drafter collaboration outperforms sequential single-drafter strategies in generation quality improvement.

These insights collectively motivate the exploration of decoupled and collaborative speculative inference on multi-nodes with heterogeneous resource.
Such cooperative strategies enable efficient adaptation to diverse task requirements while maintaining adequate resource for each phase, which is troublesome for current approaches.

\begin{figure}[t]
    \centering
    \begin{subfigure}[b]{0.49\columnwidth}
        \includegraphics[width=\linewidth]{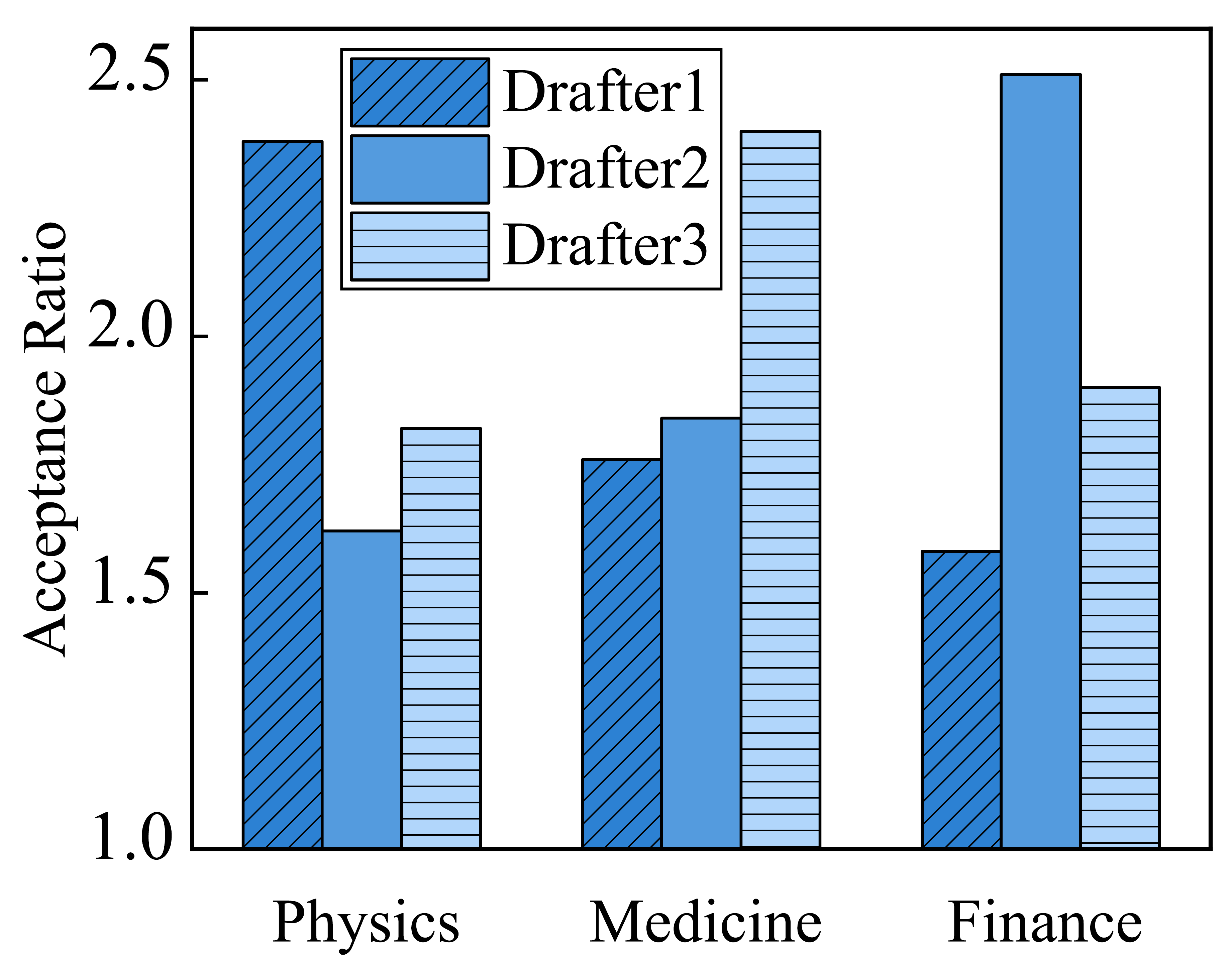}
        \caption{Differential capabilities of SSMs across various domains.} 
        \label{fig:m3}
    \end{subfigure}
    % \hfill
    \begin{subfigure}[b]{0.49\columnwidth}
        \includegraphics[width=\linewidth]{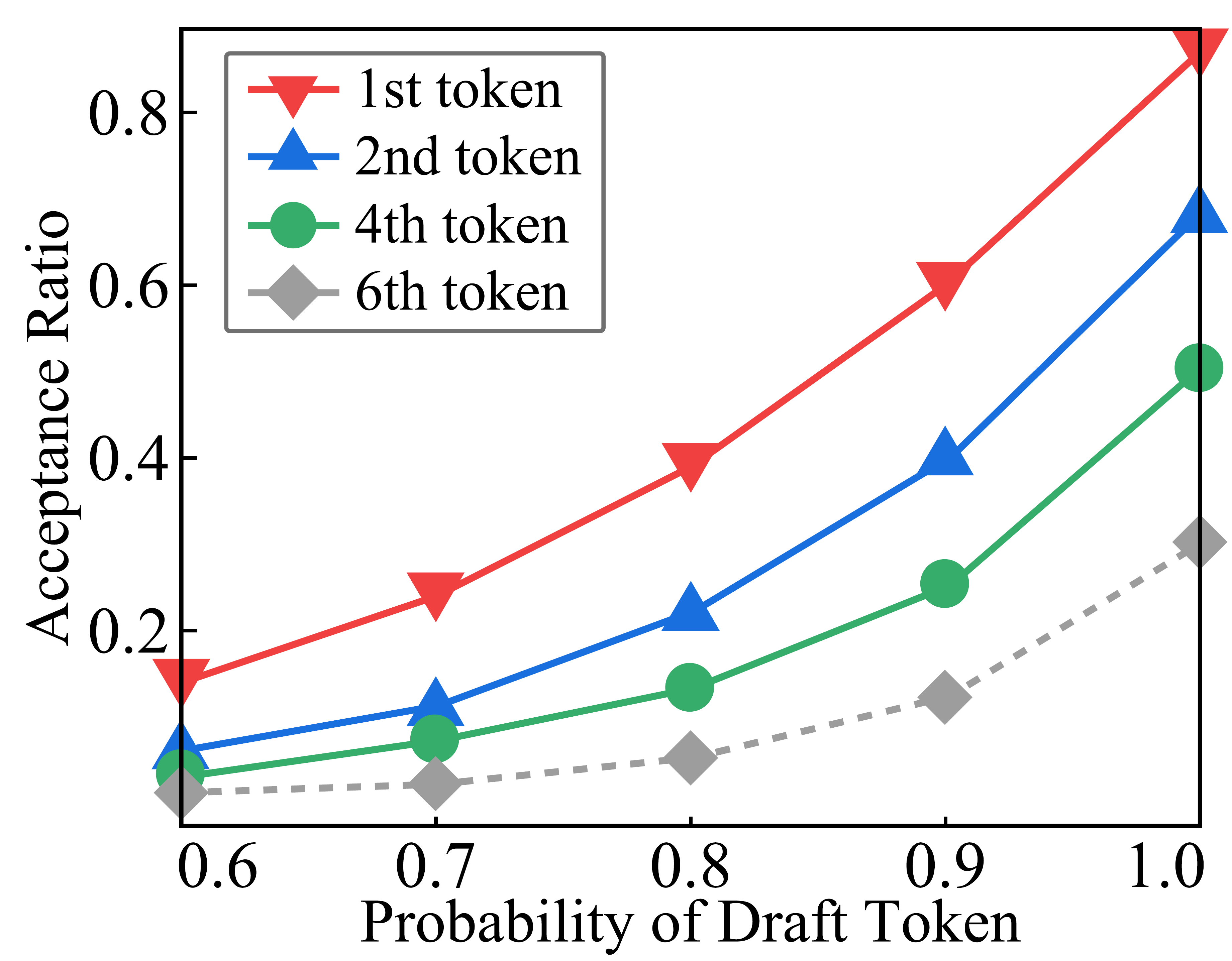}
        \caption{Acceptance ratio across various probabilities and positions.}
        \label{fig:m4}
    \end{subfigure}
    \caption{Model capabilities and token confidence in draft generation.}
    \label{fig:m34}
\end{figure}

\subsection{Opportunities of Multi-node Collaboration}

\textbf{Cross-domain Generalization with Specific Knowledge.}
With the fine-tune techniques such as knowledge distillation and domain adaptation, SSMs can specialize in different task domains and exhibit distinct capabilities.
As illustrated in \cref{fig:m3}, SSMs exhibit complementary performance across domains, with task-specific efficiency variances exceeding $2\times$.
This specialization implies that no individual drafter model excels universally, but collaborative inference can strategically integrate domain-optimized expertise.

Current static deployment strategies inadequately exploit such diversity: exhaustive deployment across all drafters introduces redundancy, while randomized selection sacrifices efficiency. 
This limitation presents an opportunity to redesign speculative inference systems through dynamic, context-aware mechanisms. 
By dynamically deploying requests to suitable drafters (\eg, activating code-optimized models for programming tasks) and intelligently aggregating high-quality tokens, systems can improve both generation quality and verification efficiency.  
The adaptive selection avoids the computational overhead in multi-node collaboration, leveraging domain expertise with low latency.

\textbf{Knowledge Integration via Confidence-Aware Fusion.}
Traditional speculative decoding uniformly weights all draft tokens, despite empirical evidence (Figure~\ref{fig:m4}) showing tokens in the top-10\% probability percentile achieve an 80\% higher acceptance ratio.
With the idea of self-speculation to leverage the knowledge behind the probability of generated tokens, it is a chance to integrate the knowledge from multiple SSMs to generate higher-quality drafts.
By aggregating outputs from parallel drafters and retaining high-confidence tokens across collaborators, we can achieve token-level expertise fusion.
Furthermore, by iteratively feedbacking selected tokens, we can form a self-correcting loop that enhances draft quality through successive refinement of low-confidence tokens. 
This confidence-aware approach offers advantages to better approximates the target LLM’s probability distribution while mitigating error propagation in drafts and improving acceptance rates.
% by leveraging the collective knowledge of multiple SSMs.

\begin{figure}[t]
    \centering
    \includegraphics[width=\columnwidth]{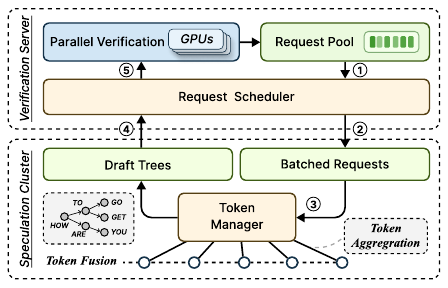}
    \caption{Overview of the CoSine architecture and workflow.}
    \label{fig:overvw}
\end{figure}

\textbf{Efficient Scheduling for Resource Utilization.}
While increasing SSM participation enhances acceptance rates, it introduces coordination overhead exhibiting superlinear growth relative to node count.
Profiling reveals two critical scheduling dimensions for optimization: (1) temporal pipelining to overlap SSM drafting with LLM verification phases, and (2) spatial load balancing that dynamically allocates batches to SSMs according to their domain specialization.
To minimize idle time and maximize resource utilization, speculative inference can implement differentiated task scheduling: memory-intensive batched drafting prioritizes high-bandwidth nodes, while compute-intensive workloads activate specialized SSMs through sparse activation patterns.
This coordinated approach ensures that each node operates at peak efficiency, reducing overall inference latency and energy consumption.
Efficient scheduling strategies can significantly enhance the performance of collaborative inference by optimizing resource utilization and minimizing idle time. % change

In summary, by transforming heterogeneous hardware constraints into distributed optimization opportunities, this paradigm enables LLM acceleration systems to transcend the sum of their parts, surpassing conventional performance ceilings. 
% achieving unprecedented efficiency and quality.
The following sections will further explore meticulously designed strategies with these opportunities of multi-node collaboration in speculative inference, aiming at achieving efficient and high-quality inference.

\section{System Design}\label{sec:design} 

% \begin{figure}[t]
%     \centering
%     \includegraphics[width=0.8\columnwidth]{chap3/pic/framework}
%     \caption{CoSine architecture overview.}
%     \label{fig:overvw}
% \end{figure}

\subsection{CoSine Overview} \label{sec:overview}

In this section, we present the architecture and workflow overview of CoSine, a collaborative LLM inference system that decouples the speculative inference process, \ie, decoding and verification, to multiple nodes for efficient inference performance.
% the main idea of CoSine xx for  scalable system
Based on the above motivations in \cref{sec:motivation}, CoSine addresses the dual challenges of terrible draft acceptance and resource efficiency by leveraging the expertise of diverse speculators and the capabilities of various nodes.
% main components
As depicted in \cref{fig:overvw}, CoSine incorporates two key components to facilitate multi-node collaboration, including the \textit{cooperative generation component} for efficient speculative decoding and the \textit{collaborative pipeline component} for adaptive workflow management and dynamic resource allocation, respectively.
% modules in CoSine 
These components are supported by two primary system modules for collaborative inference, \ie, \textit{speculation cluster} and \textit{verification server}.
Together, they enable efficient and scalable decoupled speculative inference through token-level transmission of batched requests and draft tokens.
% speculation cluster

The speculation cluster is organized as a star-topology network of consumer-grade nodes, each equipped with specialized speculators optimized for distinct linguistic patterns.
These nodes are coordinated by the cooperative generation component, which dynamically selects the most suitable speculators and conducts token fusion to generate high-quality drafts.
% verification server to generate high-quality drafts 
The verification server operates on multiple datacenter-grade GPUs, employing advanced parallelism techniques to execute efficient processing of batched draft tokens.
To bridge these modules, the collaborative pipeline component optimizes resource utilization through real-time workload balancing between draft generation and verification phases in pipeline execution.

% For seamless operation and further resource utilization between two modules, CoSine introduces the adaptive pipeline collaboration scheme to balance draft generation and verification in real-time with asynchronous execution.
% The speculation cluster is orchestrated by a central node called Token Manager, which coordinates the speculative tokens and applies token fusion of requests to generate diverse and high-quality token trees.
% With the relatively stable validation latency and substantial computational resources, CoSine can take advantage of the underutilized resources to pursue the highest level of optimization for the speculative system.
% The Request Scheduler in the server dynamically organizes the inference requests into batches and specifically schedules them in the speculation cluster for expeditious draft token generation. 

%system inference dataflow 
As illustrated in \cref{fig:overvw}, requests are processed iteratively in a fine-grained batched manner and maintained in a request pool for continuous processing. 
The system dataflow is orchestrated through a collaborative pipeline component, which begins by iteratively dispatching a continuous stream of batched requests from the request pool to the speculation cluster.
With speculative decoding of batched requests implemented (detailed in \cref{sec:ssm}), the cooperative generation component routes requests to suitable drafters based on inference and workload status.
During parallel draft generation, CoSine adopts a confidence-based token fusion method to merge draft tokens from multiple drafters, ensuring the generation of high-quality and diverse token trees.
Subsequently, the selected drafts are transmitted to the verification server under tensor and pipeline parallelism, where the transformer layers are distributed across multiple GPUs to balance the computational load. 
Once verified, requests are returned to the request pool for further scheduling until either the End-of-Sequence (\texttt{<EOS>}) token is reached or the maximum generation length is achieved.
% The integration of real-time scheme enables CoSine to deliver efficient and scalable inference capabilities suited to diverse operational demands.
%
% The design principles underlying our approach are orthogonal to existing LLM acceleration methods (\eg, quantization, pruning), suggesting additional performance improvements through collaborative inference.

% \texttt{\FSpecCluster} and \FSpecCluster{} % without speculation cluster, the inference system can convert to a traditional inference system
% The primary objective of the mechanism is to enable multiple devices to cooperatively generate draft tokens for speculative decoding, thereby alleviating the server's workload and optimizing system data flow.
\subsection{Cooperative Generation Component} \label{sec:ssm}
% structure of the speculation cluster
In this subsection, we present the architectural design and implementation of the speculation cluster for cooperative draft generation in CoSine. 
The cooperative generation component enables multiple nodes with consumer-grade GPUs to cooperatively generate draft tokens through speculative decoding, thereby improving system efficiency and scalability for batched inference requests while reducing server workload.
The speculation cluster employs a star-topology architecture with a central node for orchestration, as depicted in \cref{fig:overvw}.
This design facilitates microsecond-level coordination across nodes connected via token-based communication protocols (\eg, Ethernet, InfiniBand), supporting a wide range of collaboration scenarios from edge-based to near-cloud coordination.
Besides, the architecture enables seamless node integration and detachment while maintaining adaptive request routing and token fusion methods.
%

% request routing method with formula and algorithm  by capitalizing the strengths and mitigating the weaknesses. 
\cref{alg:ssm} details the adaptive request routing strategy that assigns inference requests to suitable drafters through multi-dimensional evaluation, including generation confidence, verification accuracy, and system status.
% token logit probability, the historical verification status and node capability metrics.
%
With node $n$ participating in the speculative inference of request $r$, it generates a $K$ length token sequence $\mathbf{X}_{n}^r = \{x_{n,1}^r, x_{n,2}^r, \ldots, x_{n,K}^r\}$, where $x_{n,i}^r$ is the $i$-th token.
The corresponding token logit probabilities $\mathbf{C}_{n}^{r} = \{P(x_{n,1}^r), P(x_{n,2}^r), \ldots, P(x_{n,K}^r)\}$ represents the generation confidence on each tokens, where $P(\cdot)\in(0,1)$ is the probability function associated with token generation.
% the importance of the confidence.
The generation confidence serves as a metric to evaluate the expertise domains of the drafters during the current inference process. 
Specifically, a drafter exhibiting high confidence in the generated draft tokens is more likely to leverage their specialized knowledge for token generation.

We also take the historical verification status into consideration, measuring the draft accuracy to ensure the quality of the draft tokens.
The verification accuracy on draft sequence $\mathbf{D}_{n}^{r} = \{d_{n,1}^r, d_{n,2}^r, \ldots, d_{n,K}^r\}$ is calculated based on the cosine similarity between the draft tokens $\mathbf{X}_{n}^r$ and the accepted tokens.
The $i$-th token draft accuracy $d_{n,i}^r$ is calculated as: 
\begin{equation}
    \label{eq:dacc}
    d_{n,i}^{r} = 
    \begin{cases}
        \cos(H(\mathbf{x}_i^r), H(x_{n,i}^r)) & \text{if } i < L_{\text{acc}} \\
        0 & \text{otherwise}
    \end{cases}, 
\end{equation}
where $L_{\text{acc}}$ represents the draft acceptance length and $\mathbf{x}_i^r$ denotes the accepted token at position $i$.
Here, $H(\cdot)$ refers to the embedding layer encoder of LLM and $\cos(\cdot)$ is the cosine similarity function.
% the importance of the token accuarcy for the draft quality, on drafter selection 
The verification accuracy reflects the historical performance of the drafters in generating high-quality tokens, thereby guiding the routing strategy to select drafters with a a demonstrated ability to produce accurate and reliable tokens.

For each request $r \in R$ in the batched requesets, we maintain a routing vector $\mathbf{M}_r = [m_1^r, \ldots, m_N^r]$, where $m_n^r \in (0,1)$ represents routing score of node $n$. 
The score combines generation confidence $c_{n,i}^r$ and verification-aligned accuracy $d_{n,i}^r$ through the normalized harmonic mean:
\begin{equation}
    \label{eq:score}
    m_{n}^r = \frac{1}{K}\sum_{i=1}^{K} \frac{c_{n,i}^{r}d_{n,i}^{r}}{c_{n,i}^{r}d_{n,i}^{r}+(1-c_{n,i}^{r})(1-d_{n,i}^{r})} \in (0,1),
\end{equation}
This formulation quantifies the synergistic effectiveness between the generation confidence and the verification accuracy.
Thus, drafters that demonstrate higher confidence and accuracy in generating tokens are assigned higher routing scores for the current request.

\begin{algorithm}[t]
    \caption{Algorithm for Cooperative Generation Component with Request Routing and Token Fusion}
    \label{alg:ssm}
    \SetKwFunction{FRoute}{RequestRouting}
    \SetKwFunction{FFuse}{TokenFusion}
    \SetKwProg{Fn}{Function}{:}{}
    
    \textbf{Input:} 
        Request $r$ in batch, Routing vector $\mathbf{M} \in (0,1)^{N}$ \\
    
    \If{$L^r_{\text{acc}} < \tau$}{\tcp{Conduct request routing with \cref{eq:exp}}
       $\text{Routing}(r) = \text{Explore}(\mathbf{M})$ \tcp{Exploration mode}
    }
    \Else{
       $\text{Routing}(r) = \text{Exploit}(\mathbf{M})$ \tcp{Exploitation mode}
    }
    
    $\text{send}(\{r, \mathbf{M}_r\}, \text{Routing}(r))$ \\\tcp{Send request to the Speculation Cluster}

    \ForEach{node $n \in N$ in parallel}{
        \ForEach{iteration $i \in [1, K]$}{
            $x_i^n \gets \text{GenerateDrafts}(x_{i-1}^n)$ \\ %\tcp{Autoregressive draft generation}         
            $x_{i-1}^n \gets \FFuse(x_{i-1}^n, \text{Routing}(r))$ \\ \tcp{Token fusion for draft generation}
        }
        Get token logit probabilities $\mathbf{c}_n$ \\
    }
    
    $\text{Tokens: } \mathbf{X} \gets \bigcup_{n=1}^N x_K^n.$ \tcp{Aggregate all draft tokens}
    $\text{Logit probabilities: } \mathbf{C} \gets \bigcup_{n=1}^N \mathbf{c}_n$ \\
    $\text{Draft tree: } \mathcal{T} \gets \text{TreeSelection}(\mathbf{X})$ \\
    Get verification accuracy $\mathbf{D}$ using \cref{eq:dacc} \\ \tcp{Draft verified in the server}
    Update routing matrix $\mathbf{M}$ using \cref{eq:conf} \\ 
    \ 
    \Fn{\FFuse{$x, \text{Routing}(r)$}}{
        Aggregate draft tokens from all nodes into $\mathbf{x}$ \\
        Retrieve token with maximum logit probabilities from $\text{Routing}(r)$: $x^* = \argmax P(\mathbf{x})$ \\ 
        $\text{send}(x^*, \text{Routing}(r))$ \\
        \Return $x^*$
    }
\end{algorithm}

The request pool maintains and iteratively updates the routing matrix $\mathbf{M} = [\mathbf{M}_1, \ldots, \mathbf{M}_R]$, which is sent to the speculation cluster alongside the batched requests.
To balance the exploration-exploitation trade-off, the routing strategy dynamically adjusts its selection policy based on the acceptance length $L_{\text{acc}}$ with exploration mode and exploitation mode.
When $L_{\text{acc}}$ falls below a predefined threshold $\tau$, the system turns to exploration mode to explore the expertise of drafters, reallocating requests to underutilized nodes. 
Otherwise, the system switches to exploitation mode, prioritizing high-performing nodes to maximize throughput.
The routing policy is formalized as follows:
% formula for selection policy with routing matrix and the exploration mode
\begin{equation}
    \label{eq:exp}
    \text{Routing}(r) = 
    \begin{cases}
        \alpha \mathcal{T}(\mathbf{M}_r) + (1 - \alpha)  \mathcal{R}(\mathbf{M}_r) & \text{if } L_{\text{acc}} < \tau, \\
        \beta  \mathcal{T}(\mathbf{M}_r) + (1 - \beta) \mathcal{R}(\mathbf{M}_r) & \text{otherwise},
    \end{cases}
\end{equation}
where $\alpha > \beta$ are the exploration coefficients, $\mathcal{T}(\cdot)$ and $\mathcal{R}(\cdot)$ are the top-scoring and random selection operators, respectively.

% token fusion method with formula / low latency network token-based communication
Furthermore, leveraging the star-topology architecture in the speculation cluster, the cooperative generation component employs a confidence-based token fusion strategy during each parallel generation iteration, as illustrated in \cref{fig:fusion}.
% maintain efficiency while enhancing the diversity of drafts  ensuring robust adaptation to diverse task requirements
In \cref{alg:ssm}, the central node aggregates draft tokens from all participating nodes and selects the token $x_i^*$ with the highest logit probability at $i$-th iteration. 
The selected token $x_i^*$ is fused into the inference sequence at the corresponding position for the subsequent generation, prioritizing confidence to optimize draft quality. 
Subsequent iterations integrate both the fused token and the historical context from all drafters, enabling the system to harness collective expertise. 
This dual dependency enhances generation diversity and quality, formalized as:
\begin{equation}
    \label{eq:conf}
    \begin{split}
        x_i^* = &\argmax_{x_i^n \in \{x_i^1, \ldots, x_i^N\}} P(x_i^n), \\
        P(x_{i+1}^n \,\big|\, x_{[1:i-1]}^n \oplus x_i^*) &\quad \text{and} \quad P(x_{i+1}^n \,\big|\, {x}_{[1:i-1]}^n \oplus x_i^n),
    \end{split}
\end{equation}
where $x_{[1:i-1]}^n$ denotes the historical context by node $n$ up to position $i-1$, and $\oplus$ represents the sequence concatenation.
This approach minimizes computational costs while operating within the constraints of limited model parameters.
Finally, the aggregated draft tokens are combined into final token trees. 
A suitable quantity and quality of tokens are then selected for the verification server using a tree-attention structure, ensuring robust adaptation to diverse task requirements.

% This cooperative mechanism allows CoSine to significantly reduce server workload while maintaining speculative verification success rates comparable to centralized approaches. % further summary 

\begin{figure}[t]
    \centering
    \includegraphics[width=\linewidth]{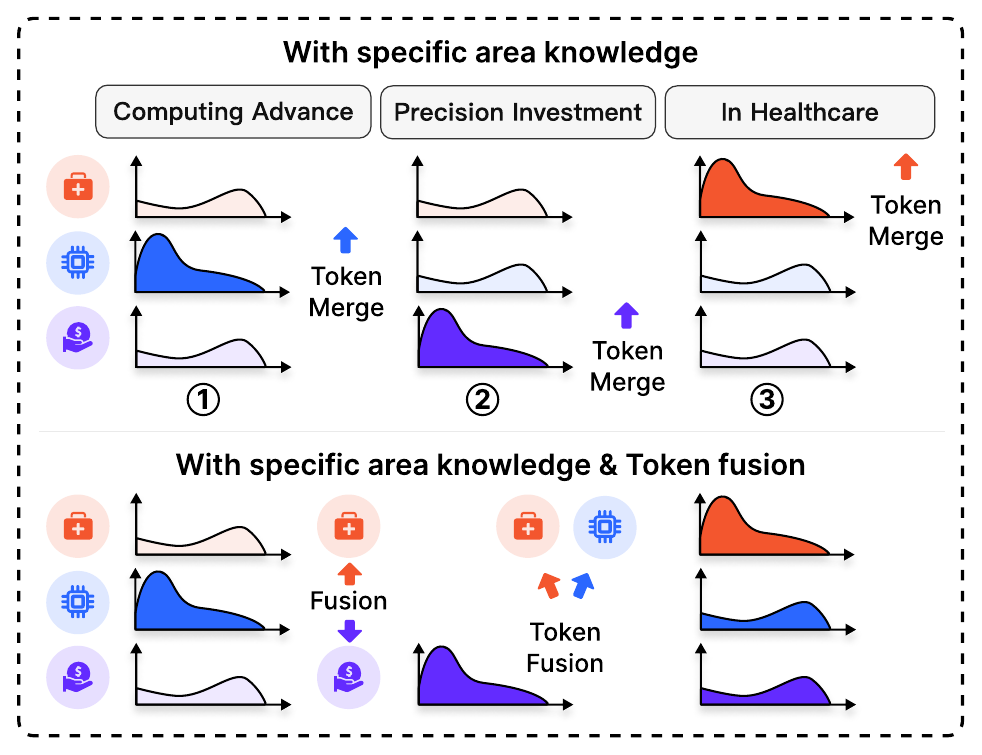}
    \caption{The token fusion process of draft generation in the speculation cluster.}
    \label{fig:fusion}
\end{figure}

\subsection{Collaborative Pipeline Component} \label{sec:llm}
This subsection details the design and operational principles of the collaborative pipeline component, which coordinates the speculative cluster and verification server in CoSine. 
% brief intro verification server and collaborative pipeline scheme (see \cref{fig:overvw})
The verification server, equipped with datacenter-grade GPUs, executes parallel verification of draft tokens while minimizing redundant computations and parameter loading overheads. 
Its architecture employs pipeline parallelism by distributing transformer layers across multiple GPUs and tensor parallelism by partitioning layer operations, enabling high-throughput token processing. 
To further optimize resource utilization, the verification server integrates continuous batching, allowing concurrent processing of multiple inference requests, and eliminating stalls between request completions. 
However, decoupling speculative inference workflows and applying phase-specific optimizations introduce idle periods and resource contention when real-time workloads deviate from projected demands.  
Such inefficiencies and intricate dependencies underscore the need for dynamic and careful coordination across heterogeneous resources.
% add some math experssion to show the idle time
 
% intro collaborative pipeline scheme with request scheduler to select requests in batch and adaptive control mechanism to adjust the speculation cluster participation dynamically
The collaborative pipeline scheme in CoSine orchestrates addresses these challenges through two core strategies with multi-node collaboration: the request scheduler and adaptive speculation mechanism.
% To further optimize the capabilities of CoSine, 
% request scheduler 
The request scheduler, illustrated in \cref{fig:overvw}, dynamically batches inference requests to balance end-to-end latency (request length) and system throughput (batch size) for further optimizing the inference capabilities.
Since batched execution latency is dominated by the longest request and batch size in the batch, the scheduler selectively groups requests from the request pool based on the current load and optimal batch size constraints.
This strategy aligns the computational contributions of the server with fluctuating workload requirements, maintaining sustained throughput and responsiveness to dynamic inference demands.

\begin{algorithm}[t]
    \caption{Collaborative Pipeline Component with Batch Assignment and Adaptive Speculation}
    \label{alg:llm}
    \SetKwFunction{FBatch}{BatchAssignment}
    \SetKwFunction{FAdaptive}{AdaptiveSpeculation}
    \SetKwFunction{FVerify}{Verify}
    \SetKwFunction{FAssign}{Assign}
    \SetKwFunction{FOutput}{Output}
    \SetKwProg{Fn}{Function}{:}{}
    
    \textbf{Input:} 
        Request pool $\mathbf{R}$, 
    \textbf{Output:} Verified tokens
    
    \While{$\mathbf{R}$ is not empty}{
        $\mathbf{B} \leftarrow$ \FBatch{$\mathbf{R}$} \tcp{Assign requests to a batch}
        $\mathbf{R} \leftarrow \mathbf{R} \setminus \mathbf{B}$ \tcp{Remove assigned requests from the pool}
        \FAssign{$\mathbf{B}$} \tcp{Assign resources for the batch}
        Send $\mathbf{B}$ and routing matrix $\mathbf{M_B}$ to the speculation cluster \\
        Get drafts $\mathcal{T}$ for each request in $\mathbf{B}$ %\tcp{Generate draft tokens}

        \ForEach{draft $t \in \mathcal{T}$ in parallel}{
            $(V_t, R_t) \leftarrow$ \FVerify{$t$} \tcp{Verify draft tokens}
            \If{Not max length and <EOS> not in $V_t$}{
                $\mathbf{R} \leftarrow \mathbf{R} \cup R_t$ \tcp{Add new requests to the pool}
                \FOutput{$V_t$} 
            }
        }
    }

    \Fn{\FBatch{$\mathbf{R}$}}{
        Model the latency of sequential speculative decoding $T_{SSM}$ and parallel verification $T_{LLM}$ \\
        Adjust draft token count $\gamma_i$ with \FAdaptive{$\mathbf{B}$, $\Gamma_{\text{max}}$} \tcp{Adaptively adjust token counts}
        Solve the linear programming problem \cref{eq:lp} to determine $\mathbf{B^*}$ % \tcp{Optimize batch assignment}
        \Return $\mathbf{B^*}$
    }
    
    \Fn{\FAdaptive{$\mathbf{B}$, $\Gamma_{\text{max}}$}}{
        \While{$\sum b_i\gamma_i > \Gamma_{\text{max}}$}{
            $j \leftarrow \argmax\gamma_i$ \tcp{Find highest probility}
            $\gamma_j \leftarrow \gamma_j - 1$ \tcp{Reduce token count for that request}
        }
    }

\end{algorithm}

% Linear Programming problem to determine xx and formulate the function Linear Programming
The request scheduler determines the batch assignment strategy for each speculative inference iteration, formalized as a vector $\mathbf{B} = \{b_1, b_2, \ldots, b_R\}$, where $b_i \in \{0,1\}$ and $R$ denotes request number in the request pool.
The batch size $b$ and critical request length $l$ can be calculated as:
\begin{equation}
    \label{eq:batch}
    b = \sum_{i=1}^{R} b_i, \quad
    l = \max_{i:b_i=1} b_il_i,
\end{equation}
where $l_i$ represents the sequence length of request $i$. 
Processing latencies for speculative decoding ($T_{\text{ssm}}(b, l, \gamma)$) and verification ($T_{\text{llm}(b, l, \Gamma)}$) are experimentally modeled as functions of $b$, $l$, and token counts $\gamma$ (draft tokens) and $\Gamma$ (verified tokens) with constraints:
\begin{equation}
    \label{eq:token}
    \begin{split}
        \Gamma &= \sum_{i=1}^{R} b_i\gamma_i, \quad \Gamma \leq \Gamma_{\text{max}}, \\
        \gamma_i &\geq 1 \quad \forall i \in \{1,\ldots,R\},
    \end{split}
\end{equation}
When processing a batch of inference requests, the end-to-end latency $T_{\text{ttl}}$ and memory consumption are constrained by the following inequalities:
\begin{equation}
    \label{eq:ttl}
    \begin{split}
    &T_{ttl} = \max(T_{ssm}(b, l, \gamma)) + T_{llm}(b, l, \Gamma), \\
    &T_{ttl} \leq T_{\text{max}}, \\
    &\sum_{i=1}^{R} b_im_i \leq M_{\text{max}}, 
    \end{split}
\end{equation}
where $m_i$ denotes per-request memory footprint of request $i$, $T_{\text{max}}$ and $M_{\text{max}}$ represent the maximum allowable latency and memory consumption, respectively.
The primary optimization objective of the request scheduler is to co-optimize throughput and latency for batched inference requests. 
This is formulated as a linear programming problem as follows:
\begin{equation}
    \label{eq:lp}
    \begin{split}
    \mathbf{B}^* &= \argmin_{\mathbf{B}} \left( \frac{T_{\text{ttl}}}{b} + \lambda\Gamma \right), \\
    \st &\quad \cref{eq:token}, \cref{eq:ttl}
    \end{split}
\end{equation}
where $\lambda$ is a weighting factor that balances the trade-off between maximizing the verified token length and minimizing latency. 
By adjusting $\lambda$, the system can prioritize either throughput or responsiveness based on operational requirements.

% adaptive speculation to minimize idle time in pipeline
According to Amdahl's Law, the potential improvement in system throughput is fundamentally constrained by the latency of sequential execution.
The adaptive speculation mechanism is introduced to adapt the draft generation phase to align with the verification timeline, using idle time in pipeline while improving the generation diversity.
As illustrated in \cref{alg:llm}, the adaptive speculation mechanism dynamically adjusts the participation of the speculation cluster based on the verification server's processing status.
When the verification server is idle, the speculation cluster increases the number of participating nodes to generate draft tokens, minimizing idle periods and enhancing system throughput.
Conversely, when the verification server is overloaded, the speculation cluster reduces the number of participating nodes to alleviate resource contention and maintain verification throughput.
This adaptive control mechanism ensures that the speculative inference system operates at peak efficiency, balancing draft generation and verification workloads to maximize resource utilization and system throughput.

\section{Implementation} \label{sec:imp}
% \textbf{Implementation details.}
In order to evaluate the performance of CoSine, we utilize Python to implement a speculative inference system on two physical prototypes, \ie, the NVIDIA A100 GPUs (80GB) and NVIDIA RTX 2080Ti GPUs.
The Python-based prototype (4.2k LoC) integrates PyTorch 2.1 and DeepSpeed 0.12, with four key innovations: (1) A star-topology speculation cluster employing specialized drafters (TinyLlama-1.1B, Phi-2), (2) A verification server with hybrid parallelism, (3) Confidence-aware token fusion, and (4) Dynamic pipeline orchestration. 

The speculation cluster operates on Ubuntu 22.04 with Docker-containerized drafters (--memory="8g" --cpus=4 limits), each hosting distinct model variants optimized for specific linguistic patterns through knowledge distillation. 
Our cooperative generation engine dynamically routes requests using a lightweight linear programming solver (0.1ms decision latency) that analyzes lexical richness and syntactic complexity via PyTorch Geometric, selecting 2-3 drafters per request. 
The token fusion process combines draft tokens, leveraging confidence scores and historical verification accuracy to ensure high-quality drafts.

%low latency sub-millsec
The verification server occurs on A100 GPUs through DeepSpeed-optimized tensor/pipeline parallelism, sharding Llama-2-70B across 4 GPUs with 4-stage pipelining (16 microbatches).  
We modify HuggingFace's generate() with CUDA-accelerated rejection sampling, achieving 1.7× faster verification than sequential decoding.

\section{PERFORMANCE EVALUATION}\label{sec:evaluation}

In this section, we address the following key research questions:
\begin{itemize}
\item How does CoSine's performance scale with varying batch sizes compared to state-of-the-art baselines across different LLM pairs? \cref{sec:exp2}
\item How does CoSine achieve cost efficiency and maintain performance in online services with fluctuating request arrival rates? \cref{sec:exp3}
\item What factors contribute to CoSine's superior performance compared to existing methods? \cref{sec:exp4}
\end{itemize}

\subsection{Experiment Setup}

\textbf{System Configuration.}
We evaluate CoSine’s performance across two hardware configurations: (1) a high-performance server and (2) a heterogeneous GPU cluster, as detailed in \cref{tab:hardware}.
The server configuration comprises an AMAX deep learning workstation with four NVIDIA A100 (80GB) GPUs interconnected via NVLink, optimized for parallelized LLM inference. 
% The server is performed for Parallelized LLM inference validation 
The heterogeneous GPU cluster consists of 16 consumer-grade GPUs (8 NVIDIA RTX 2080Ti and 8 RTX 3090 GPUs) as individual nodes, interconnected via a 100Mbps Ethernet network.
% This configuration achieves a 47 $\times$ reduction in rental costs compared to datacenter-grade GPUs. 
Both hardwares are collaboratively utilized for speculative inference tasks connected with a 10Gbps Ethernet network with sub-1ms latency.

\begin{table}[t]
    \centering
    % \small
    \caption{Performance and cost comparison of high-performance server and consumer-grade GPUs.}
    \label{tab:hardware}
    
    \begin{tabular}{lccc}
        \toprule
        \textbf{Metrics}        & \textbf{2080Ti} & \textbf{3090} & \textbf{A100} \\
        \midrule
        FPLOPS (FP16)           & 107.6        & 285          & 5144   \\
        Bandwidth (GB/s)        & 616          & 936          & 2,039  \\
        SSM Speed (tokens/s)    & 350          & 450          & 9,500  \\
        LLM Speed (tokens/s)    & \textit{OOM} & \textit{OOM} & 7.13   \\
        Rent Cost (\$/hr)       & 0.12         & 0.22         & 5.67   \\
        Deploy Cost (\$)        & 200          & 1,000        & 60,000 \\
        \bottomrule
    \end{tabular}
    \begin{tablenotes}
\small
\item *\textit{OOM}: Out of memory
\end{tablenotes}
\end{table}

\textbf{Tested Prompts.} 
Our evaluation employs prompts from five domain-specific datasets: PIQA (physics) \cite{Bisk2020}, MedQA (medicine) \cite{jin2020disease}, FIQA (finance) \cite{thakur2021beir}, Alpaca (instruction following) \cite{alpaca}, and OASST2 (conversational alignment). 
We randomly average sample 8192 prompts across these datasets, preserving their original proportionality to simulate real-world application scenarios. 
This cross-domain selection introduces challenging speculation conditions due to diverse linguistic structures and domain-specific constraints.

\textbf{Model Settings.} %three orders of magnitude
We evaluate two LLM pairs with distinct parameter scales.
The \textbf{LLaMA pair} consists of DeepSeek-R1-Distill-Llama-70B \cite{touvron2023llama} (target model) and LLaMA68M \cite{miao2024specinfer} (drafter), with a parameter ratio difference on the order of millions (evaluated with 2080Ti GPUs in cluster).
The \textbf{Qwen pair} includes DeepSeek-R1-Distill-Qwen-32B (target model) and Qwen2.5-0.5B \cite{yang2024qwen2} (drafter), with a parameter ratio difference on the order of hundreds (evaluated with 3090 GPUs in cluster).
As shown in \cref{tab:dataset}, drafters (\#1–\#6) exhibit task-dependent expertise due to domain-specialized fine-tuning on the aforementioned datasets, resulting in varied draft acceptance rates (from 1.73 to 3.20). 
All models use float16 precision for parameters and activations, with weight consistency maintained across all SSMs and LLMs.
%  to ensure fair performance comparisons. 

\textbf{Baselines.} % costs of deploying GPUs or renting GPUs from cloud providers. 
% To the best of our knowledge, there are no collaborative speculative inference systems applicable to our scenario.
We compare CoSine against the following baselines to evaluate its performance in draft generation and multi-node collaboration:

\begin{itemize} %intro decouiple status
    \item \textbf{vLLM} \cite{yuan2024llm}: A LLM inference framework that implements continuous batching with distributed execution across GPUs, as the baseline for server-side execution.

    \item \textbf{Vanilla Speculative Inference (Vanilla)} \cite{leviathan2023fast}: An extension of vLLM that employs a single draft model for speculative decoding and verification, executed in a coupled sequential manner.

    \item \textbf{PipeInfer} \cite{butler2024pipeinfer}: A speculative inference system featuring a decoupled pipeline architecture with asynchronous draft generation and early-exit mechanisms.  

    \item \textbf{SpecInfer} \cite{miao2024specinfer}: A speculative inference system that utilizes multiple drafters to generate tree-structured drafts while maintaining coupled synchronization for collective candidate evaluation.  
\end{itemize}

\begin{figure*}[t]
    \centering
    \newcounter{tempfigcnt}\setcounter{tempfigcnt}{\value{figure}}
    \begin{subfigure}{0.55\textwidth}
        \includegraphics[width=\textwidth]{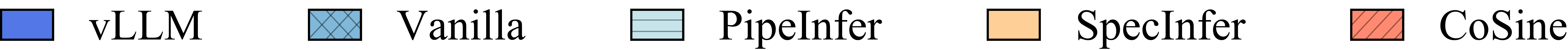}
    \end{subfigure}
    \vspace{-2mm}
    \addtocounter{figure}{-1}\setcounter{figure}{\value{tempfigcnt}}
\end{figure*}
\begin{figure*}[t]
    \centering
    \begin{subfigure}[b]{0.24\textwidth}
        \includegraphics[width=\textwidth]{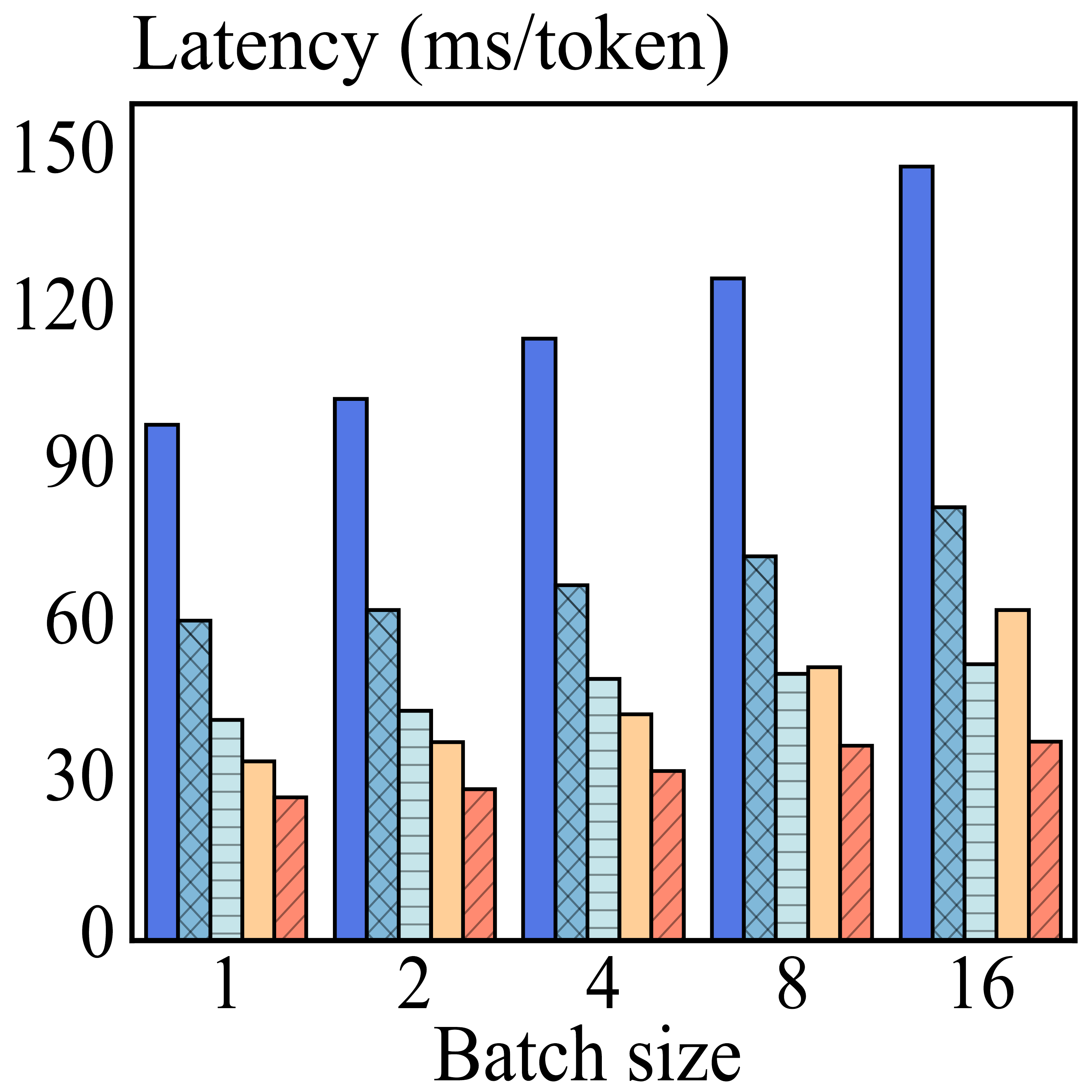}
        \caption{Latency of LLaMa pair.}
        \label{fig:ltc1}
    \end{subfigure}
    \hfill
    \begin{subfigure}[b]{0.24\textwidth}
        \includegraphics[width=\textwidth]{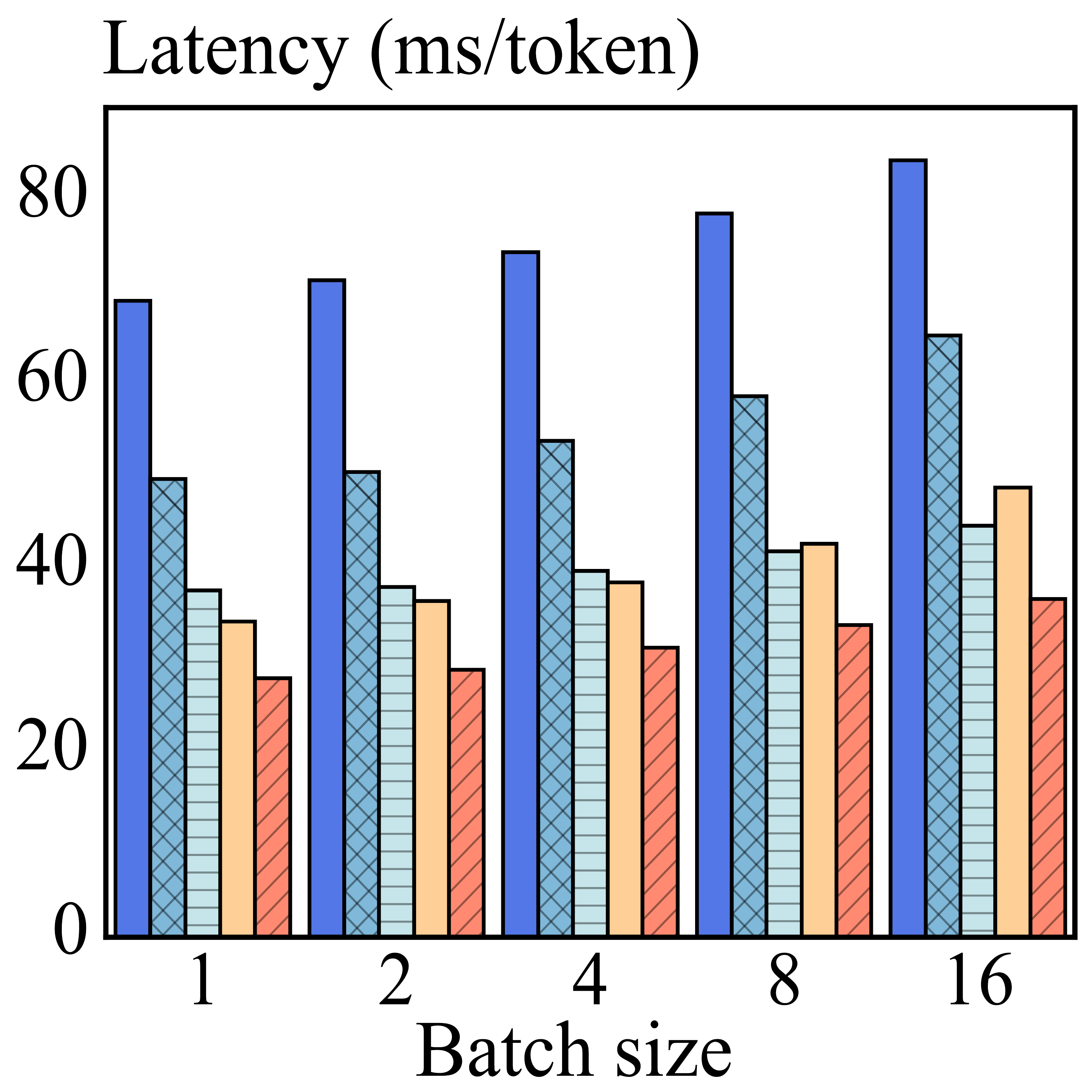}
        \caption{Latency of Qwen pair.}
        \label{fig:ltc2}
    \end{subfigure}
    \hfill
    \begin{subfigure}[b]{0.24\textwidth}
        \includegraphics[width=\textwidth]{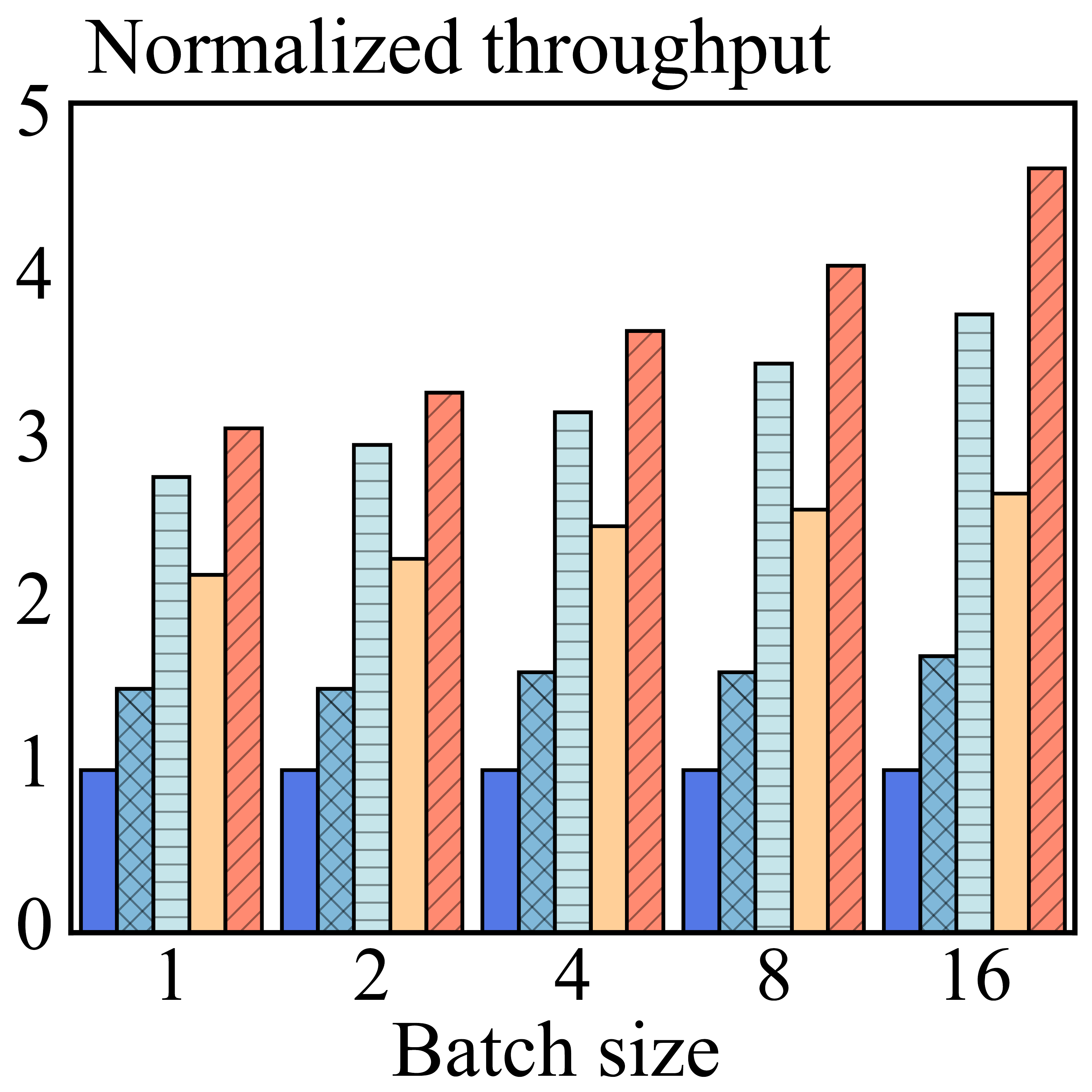}
        \caption{Throughput of LLaMa pair.}
        \label{fig:tp1}
    \end{subfigure}
    \hfill
    \begin{subfigure}[b]{0.24\textwidth}
        \includegraphics[width=\textwidth]{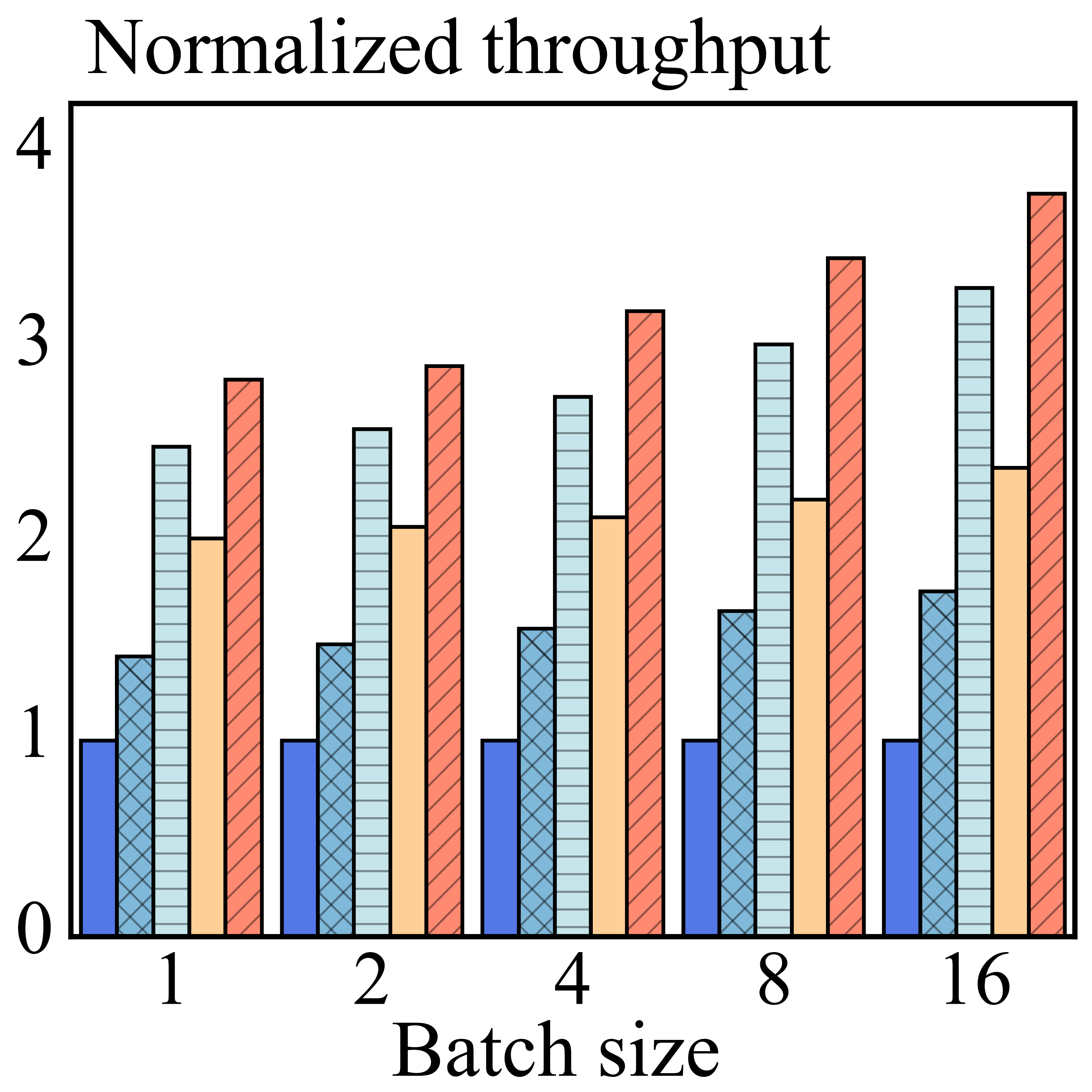}
        \caption{Throughput of Qwen pair.}
        \label{fig:tp2}
    \end{subfigure}
    
    \caption{Offline serving latency and throughput of CoSine against baselines on LLaMA pair and Qwen pair.}
    \label{fig:tp}
\end{figure*}

\textbf{Evaluation Metrics.}
We quantify LLM inference performance using the following metrics:

\begin{itemize}
    \item \textbf{End-to-End Latency} (ms/token): The average time from inference initiation to final token generation, measuring system responsiveness.
    % \item \textbf{Inter-token latency} (ms): Average temporal gap between consecutive token generations, isolating inter-token processing delays.  
    \item \textbf{Throughput} (token/s): The total number of tokens processed per second across concurrent requests, reflecting the scalability in batch processing.
    
    \item \textbf{Cost efficiency} (costs/token): The total operational costs normalized by the number of tokens generated, evaluating resource utilization efficiency.
    
\end{itemize}

% This dual-metric method enables holistic evaluation of both interactive responsiveness (via latency) and high-load capacity (via throughput), with all measurements normalized per GPU to facilitate cross-system comparisons.

\textbf{Experiments Settings.} 
% We scale the average arrival rate to 75% of the cluster’s peak throughput to avoid bursts of requests leading to OOM in the system. 
All experiments use fixed-length input sequences of 256 tokens and generate outputs of 128 tokens using greedy sampling for both draft generation and verification.
We evaluate CoSine and baselines under two settings to measure LLM serving performance.
For \textbf{offline serving}, we measure latency and throughput with fixed batch sizes ranging from 1 to 16 tokens. 
Results are reported as mean values from 10 independent trials, with 95\% confidence intervals.
Second, for \textbf{online serving}, we warm up the system for 1 minute and conduct tests over 2 hours.
We evaluate latency and cost efficiency under varying request arrival rates to simulate real-world LLM request scenarios.
To ensure fair cost comparisons, performance metrics are computation-normalized to eliminate biases arising from hardware scaling and heterogeneous GPU participation in the system.

\begin{table}[t]
    \centering
    \small
    \caption{Acceptance ratio comparison of different drafters on various datasets.}
    \label{tab:dataset}
    
    \begin{tabular}{lcccccc}
        \toprule
        \textbf{Datasets} & \textbf{\#1} & \textbf{\#2} & \textbf{\#3} & \textbf{\#4} & \textbf{\#5} & \textbf{\#6} \\
        \midrule
        PIQA   & \textbf{2.86} & 1.82  & 1.74  & 1.80  & 1.91  & 2.20  \\
        MedQA  & 1.72          & \textbf{3.13} & 1.85 & 2.01  & 1.84  & 2.05  \\
        FIQA   & 1.69          & 1.73  & \textbf{2.95} & 1.96  & 2.28  & 2.13  \\
        Alpaca & 1.83          & 1.95  & 2.04  & \textbf{2.86} & 2.13  & 1.82  \\
        OASST2 & 2.24          & 1.84  & 1.87  & 1.94  & \textbf{3.20} & 2.08  \\
        \bottomrule
    \end{tabular}
\end{table}

\subsection{Performance of Offline Serving} \label{sec:exp2}
% We first present the end-to-end inference latency and system throughput of CoSine and baselines on LLaMa pair and Qwen pair in \cref{fig:ltc} and \cref{fig:tp} across various batch sizes.
%

\textbf{Inference Latency.} %2figs
\Cref{fig:ltc1} and \cref{fig:ltc2} illustrate the inference latency of CoSine compared to baseline methods across the LLaMA pair and Qwen pair in offline serving scenarios.
The results demonstrate that CoSine consistently outperforms the baselines across batch sizes, model sizes and task domains, achieving significant latency reductions of up to 27.1\% for the LLaMA pair and 20.5\% for the Qwen pair.

For the LLaMA pair, which features parameter ratios on the order of millions, CoSine reduces inference latency by 17.9\%–27.1\% relative to the strongest baselines (\eg, SpecInfer for small batches and PipeInfer for large batches). 
Notably, for the Qwen pair, which has smaller parameter ratios, CoSine maintains a competitive edge, achieving latency reductions of 15.2\%–20.5\% compared to the most effective baseline. 
These improvements are primarily attributed to CoSine's specialized request routing and token fusion mechanisms, which leverage the collaborative efforts of multiple drafters to generate high-quality drafts and adapt seamlessly to diverse task domains.

As batch sizes scale from $B=1$ to $B=16$,  CoSine demonstrates exceptional stability, exhibiting latency variations of only 23\%. This is significantly lower than the 43\% variation observed in baseline methods.
%
% For smaller batch sizes, CoSine capitalizes on its high acceptance rate to reduce verification iterations, achieving latency reductions of 58.4\% and 50.2\% over Vanilla, and 15.2\% and 20.5\% over the best-performing baseline, SpecInfer, for the LLaMA and Qwen model pairs, respectively.
%
For larger batch sizes, CoSine leverages its batch scheduling mechanism to group requests effectively, achieving latency reductions of 22.4\% and 17.9\% over the best-performing baseline, PipeInfer, for the LLaMA and Qwen pairs.
% which dynamically optimizes request deployment across multiple SSMs (small-scale models). 
% his mechanism dynamically optimizes request deployment across multiple small-scale models (SSMs), in contrast to baseline approaches that rely on static algorithms lacking such context-aware adaptability. 
These capabilities enable CoSine to achieve superior workload adaptation and scalability in real-world inference scenarios.

\textbf{System Throughput.} %2figs
\Cref{fig:tp1} and \cref{fig:tp2} compare the normalized throughput of CoSine with baseline methods, demonstrating CoSine's high resource utilization and scalability across diverse scenarios.
For consistency and better comparison, throughput values are normalized to vLLM's throughput at each batch size (set as 1.0).

By enhancing the adaptive pipeline collaboration scheme, CoSine achieves a balanced workflow between draft generation and verification, minimizing resource inefficiency and idle periods. 
For the LLaMA pair, CoSine achieves throughput improvements of $1.31\times$ to $1.62\times$ over SpecInfer and $1.24\times$ to $1.46\times$ over PipeInfer as batch sizes increase from $B=1$ to $B=16$. 
%
% Similarly, for the Qwen pair, CoSine attains gains of $1.24\times$ to $1.52\times$ over SpecInfer and $1.17\times$ to $1.35\times$ over PipeInfer. 
%
CoSine further exhibits exceptional scalability, with larger batch sizes yielding better normalized throughput. 
It outperforms vLLM by $3.15\times$ to $4.71\times$ for the LLaMA pair and $2.84\times$ to $3.79\times$ for the Qwen pair. 
% 's decoupled architecture and its optimization of collaboration, which
This accelerated scaling is driven by the reduction of the memory bottleneck of speculative decoding and the computing bottleneck of verification, thereby providing a higher inference ceiling.

% CoSine’s stable latency profile minimizes pipeline stalls, and (2) its idle time reduction mechanism enhances hardware utilization. 
% These findings highlight CoSine’s ability to leverage speculative parallelism while maintaining resource efficiency across diverse workloads. 

\subsection{Performance of Online Services}\label{sec:exp3}
We evaluate the Online Services performance of CoSine by comparing the token generation latency of CoSine with baselines across LLaMA pair.
As shown in \cref{fig:online}, the request arrival rate is set as three modes (low, high and fluctuated) to simulate different LLM request service scenarios, across 240 miuntes. 
CoSine consistently outperforms in fast request precessing with the baselines across all request arrival rates.

In the low arrival rate scenario, CoSine achieves 1.2x-1.5x lower latency compared to the strongest baseline, SpecInfer.
In the high arrival rate scenario, CoSine reduces latency by 1.3x-1.6x relative to SpecInfer.
In the fluctuated arrival rate scenario, CoSine maintains a 1.2x-1.4x latency reduction compared to SpecInfer.
The results demonstrate that CoSine is capable of efficiently processing LLM requests in online services, providing a significant performance advantage over existing methods.

% Lower reliance on premium cloud instances: By offloading speculative tasks to low-tier cloud instances (e.g., AWS G4, T4-based VMs) and reserving high-cost instances (e.g., A100/H100) only for validation, operational costs are significantly reduced.

% Enhanced resource utilization: Idle or underutilized low-end devices (e.g., edge nodes, legacy GPUs) are repurposed for speculative workloads, improving infrastructure ROI without additional hardware investments.

\begin{figure}[t]
    \centering
    \includegraphics[width=\columnwidth]{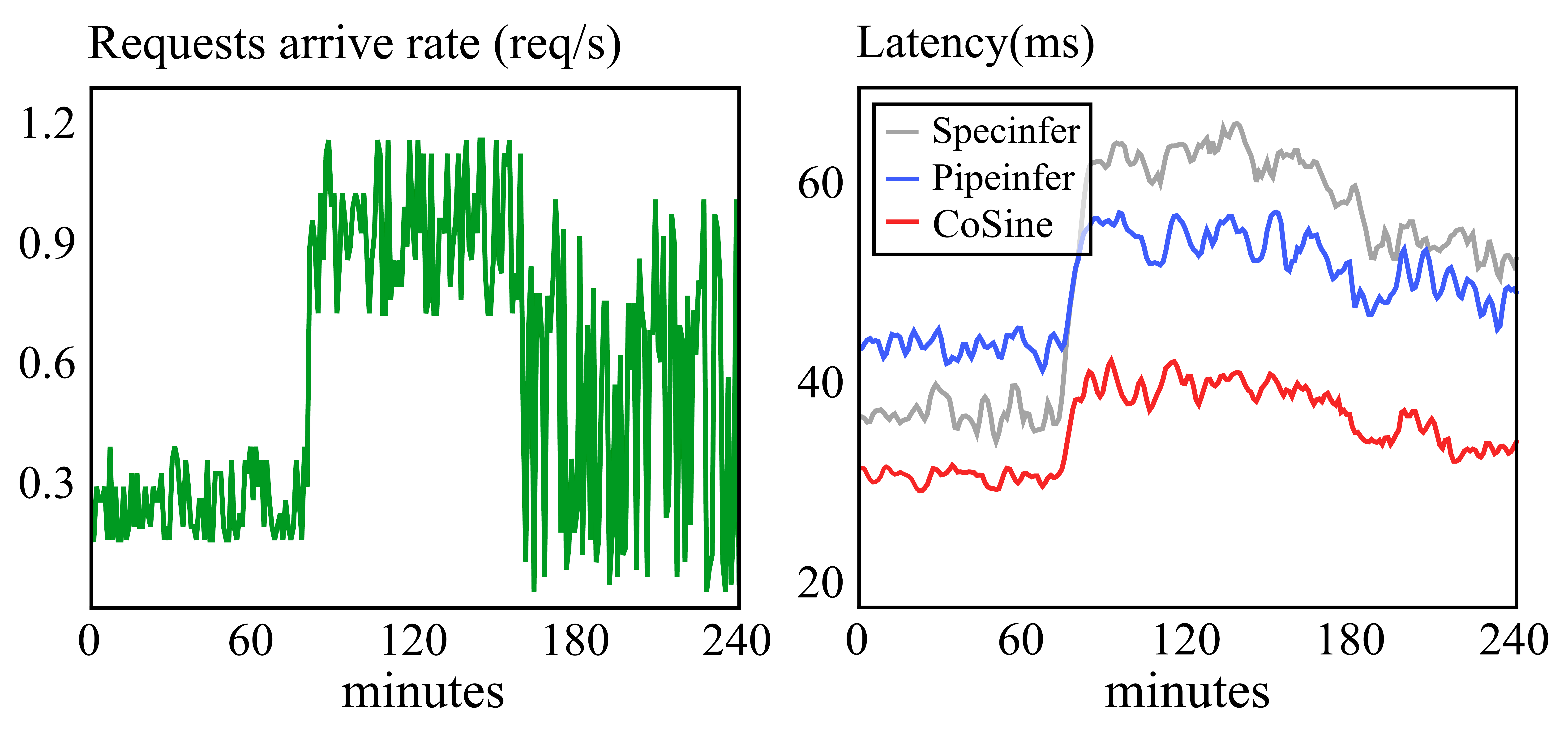}
    \caption{Online serving latency of CoSine and baselines with LLaMA pair.}
    \label{fig:online}
\end{figure}

\begin{table}[t]
\centering
\small
\caption{Cost Efficiency Comparison of CoSine and Baselines}
\label{tab:cost}
\begin{tabular}{@{}lcccccc@{}}
\toprule
\multirow{2}{*}{{Mode}} & \multicolumn{2}{c}{\textbf{SpecInfer}} & \multicolumn{2}{c}{\textbf{PipeInfer}} & \multicolumn{2}{c}{\textbf{CoSine}} \\
\cmidrule(lr){2-3} \cmidrule(lr){4-5} \cmidrule(lr){6-7}
 & LLM & Qwen & LLM & Qwen & LLM & Qwen \\ 
\midrule
Low         &43.34\%	&47.79\%	&33.23\%	&38.57\%	&\textbf{29.98\%}	&\textbf{34.34\%} \\
High       &36.87\%   &41.8\%	&26.18\%	&30.19\%	    &\textbf{21.18\%}	&\textbf{26.36\%} \\
{Volatile}  &38.01\%	&42.87\%	&28.61\%	&33.07\%	    &\textbf{25.71\%}	&\textbf{30.63\%} \\
\bottomrule
\end{tabular}
\end{table}

\subsection{Ablation Study} \label{sec:exp4}

\begin{figure}[t]
    \centering
    \begin{subfigure}[b]{0.235\textwidth}
        \includegraphics[width=\textwidth]{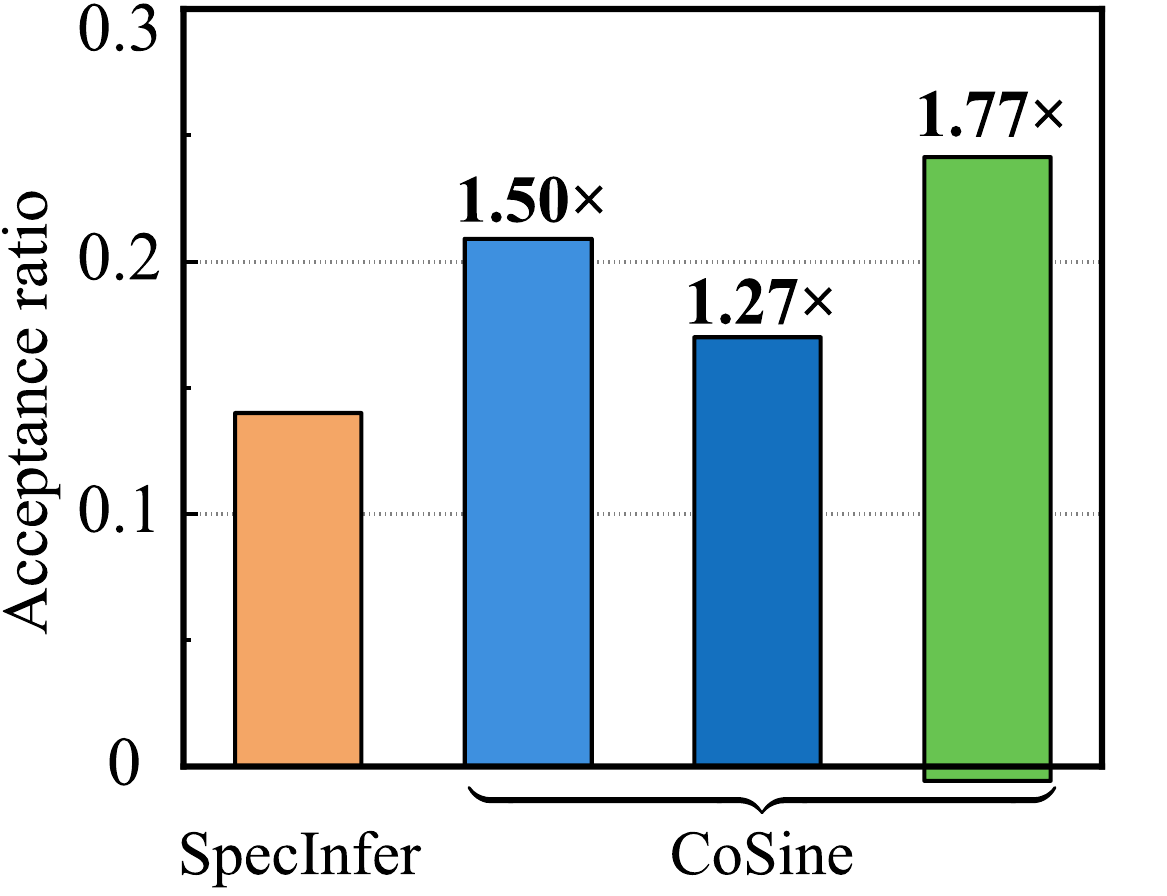}
        \caption{1 cooperative node.}
        \label{fig:abla1}
        \vspace{0.1in}
    \end{subfigure}
    \begin{subfigure}[b]{0.235\textwidth}
        \includegraphics[width=\textwidth]{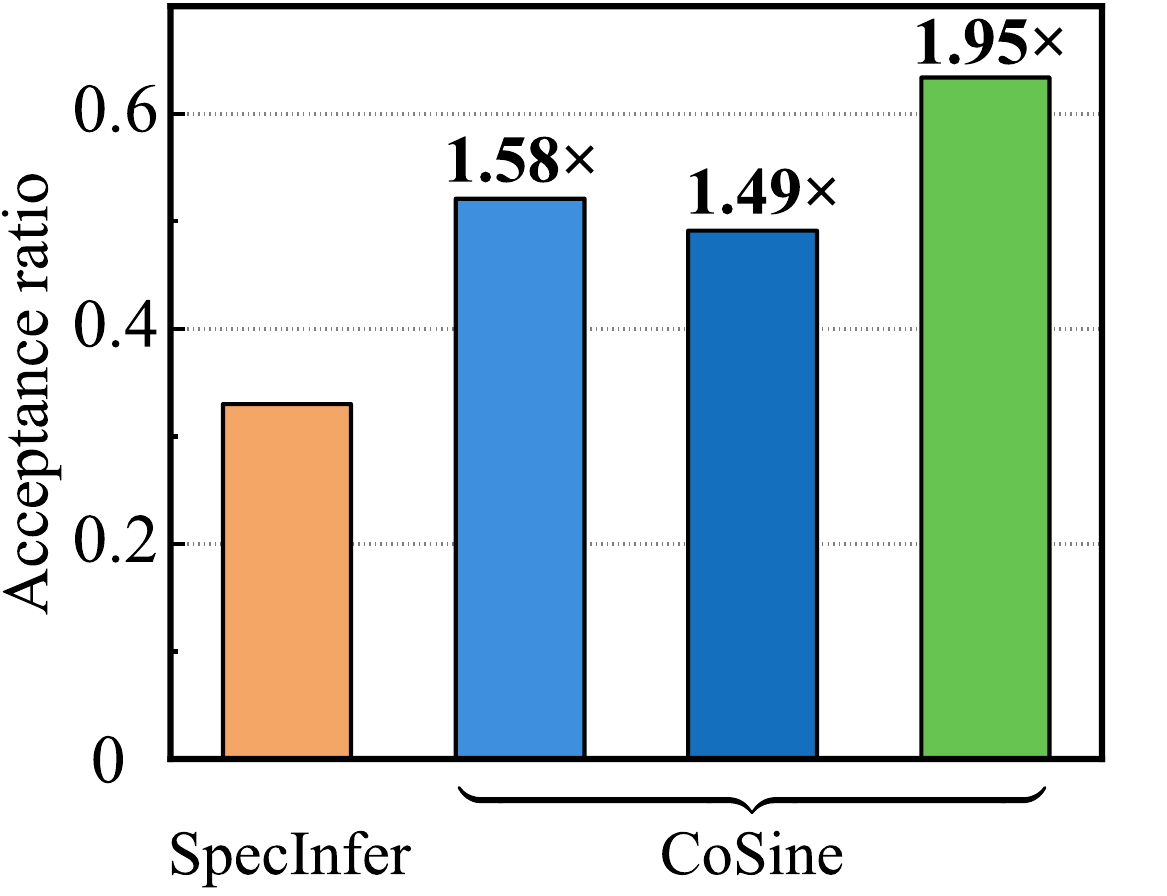}
        \caption{2 cooperative nodes.}
        \label{fig:abla2}
        \vspace{0.1in}
    \end{subfigure}
    \begin{subfigure}[b]{0.235\textwidth}
        \includegraphics[width=\textwidth]{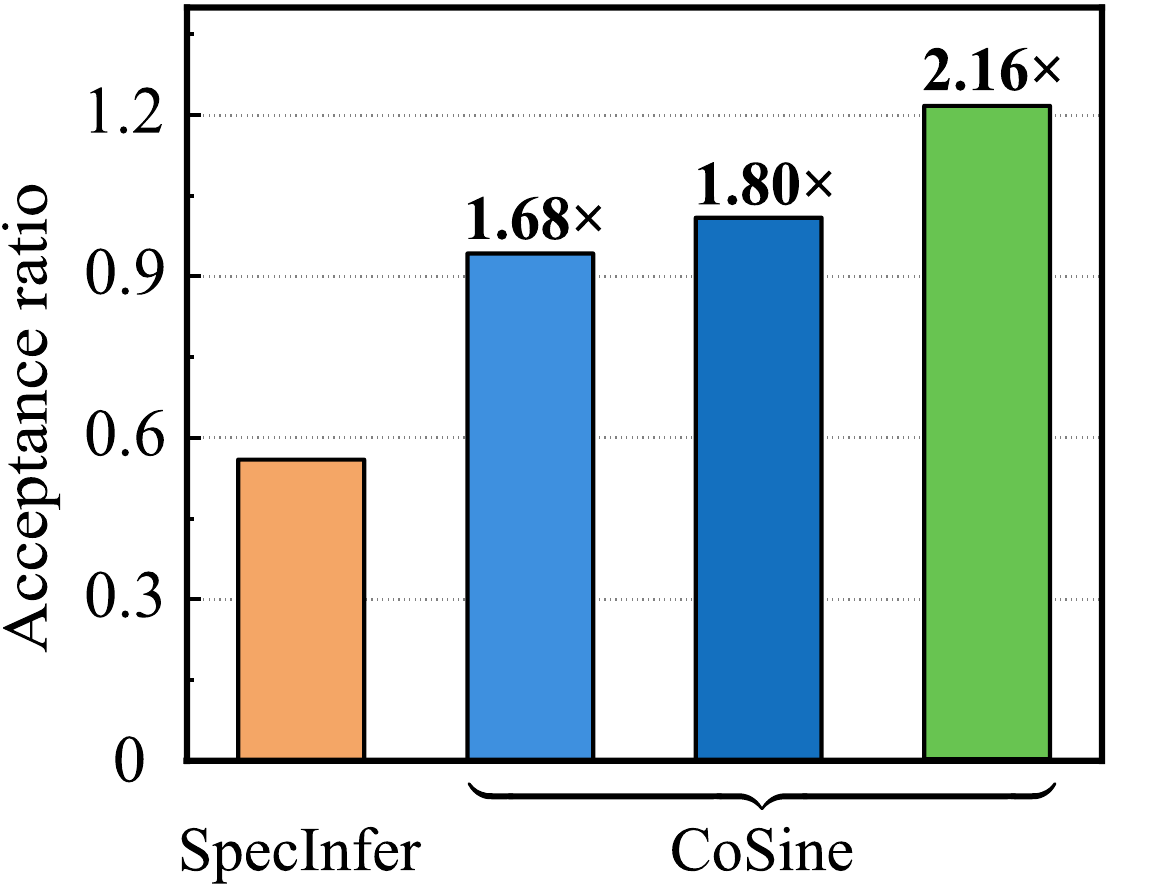}
        \caption{4 cooperative nodes.}
        \label{fig:abla3}
    \end{subfigure}
    \begin{subfigure}[b]{0.235\textwidth}
        \includegraphics[width=\textwidth]{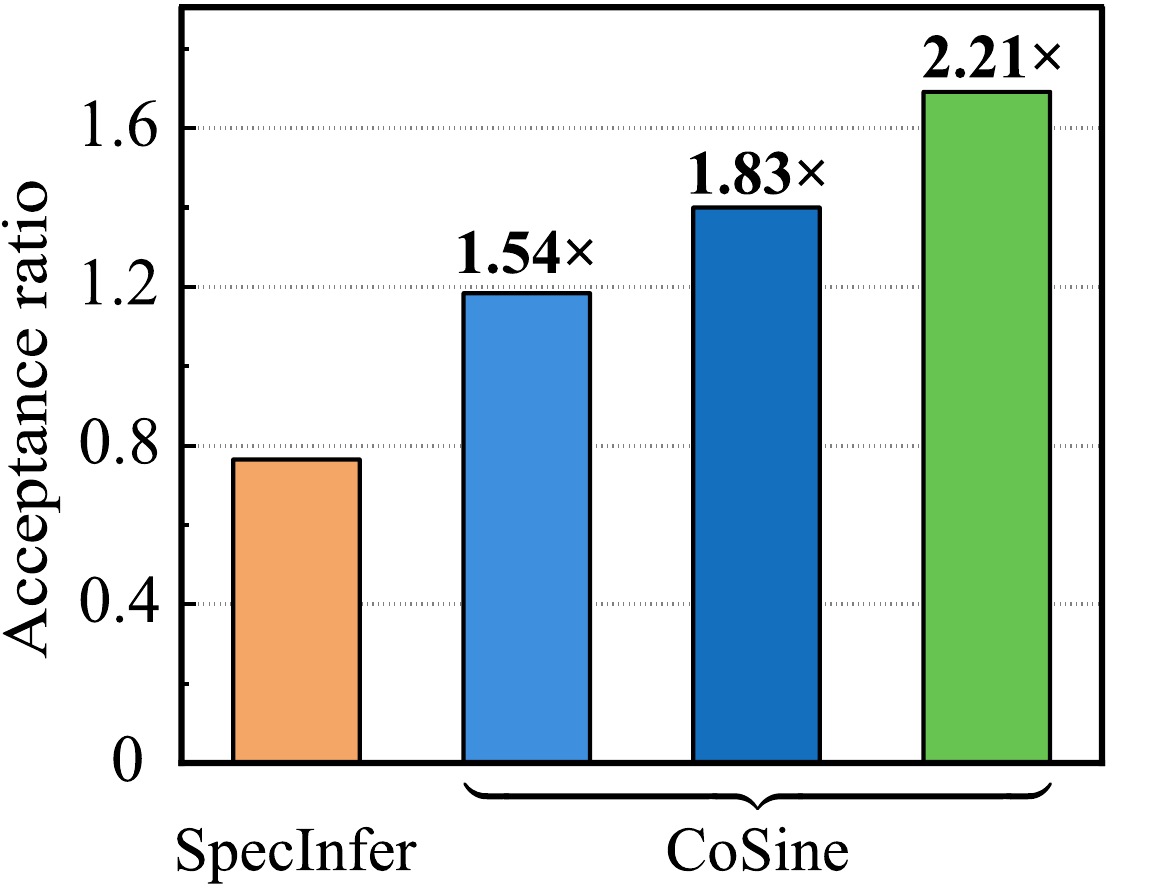}
        \caption{8 cooperative nodes.}
        \label{fig:abla4}
    \end{subfigure}    
    \caption{The acceptance ratio improvement with different numbers of cooperative nodes.}
    \label{fig:abla}
\end{figure}

To validate the efficacy of CoSine, we conduct an ablation study comparing its full implementation against variants with key components removed, as well as the SpecInfer baseline. The results (\cref{fig:abla}) demonstrate how CoSine’s cooperative generation mechanism and collaborative pipeline scheme collectively enable scalable performance gains across device scales.

\textbf{Component Effectiveness.} 
Without cooperative generation mechanism exhibits 29-33\% lower throughput than full CoSine, with the gap widening at larger scales (8 devices: 1.18 \vs 1.72). This degradation stems from the inability to dynamically route requests to specialized drafters in the speculation cluster. Without intelligent drafter selection based on linguistic patterns and workload status, nodes generate less relevant drafts that require more verification iterations, particularly evident in complex multi-node scenarios where heterogeneous requests amplify the need for specialization.

Disabling confidence-based token fusion (without token funsion) reduces throughput by 17-34\%, highlighting the importance of synthesizing outputs. The fusion mechanism compensates for individual drafter limitations by merging complementary token candidates, creating higher-quality draft trees. At 8 devices, the 1.13→1.72 improvement demonstrates how fusion prevents quality saturation as parallel drafters increase—without aggregation, added nodes yield diminishing returns due to redundant low-confidence tokens.

\begin{figure}[t]
    \centering
    \newcounter{tempfigcnt1}\setcounter{tempfigcnt1}{\value{figure}}
    \begin{subfigure}{0.43\textwidth}
        \includegraphics[width=\textwidth]{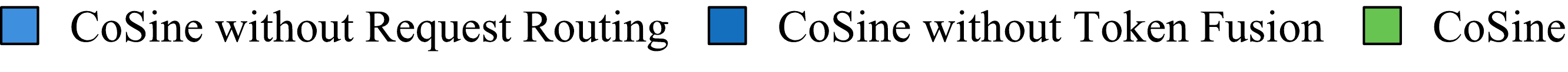}
    \end{subfigure}
    % \vspace{-0.3in}
    \addtocounter{figure}{-1}\setcounter{figure}{\value{tempfigcnt}}
\end{figure}

\section{Conclusion}\label{sec:discussion}
In this paper, we introduce a framework for efficient LLM inference system called CoSine that utilizes collaborative resources for speculative inference. 
CoSine refines the verification mechanism for direct ensemble sampling and introduces an alternate proposal framework to further boost efficiency. 
We demonstrate the effectiveness of CoSine through both theoretical analysis and empirical validation.
Our results show that CoSine achieves significant improvements in inference latency, throughput, and cost efficiency compared to state-of-the-art LLM inference systems.

\balance
\bibliographystyle{unsrt}
%unsrt plain ACM-Reference-Format
\bibliography{ref.bib} 

\end{document}